\documentclass[journal, 10pt]{IEEEtran}

\usepackage{cite}
\usepackage{amsmath,amssymb,amsfonts}
\usepackage{algorithmic}
\usepackage{graphicx}
\usepackage{subfigure}
\usepackage{textcomp}
\usepackage{xcolor}
\usepackage{url}
\usepackage{bbm}
\usepackage{booktabs}

\ifCLASSINFOpdf
\else
\fi
\begin{document}

\title{Sensing Integrated DFT-Spread OFDM Waveform and Deep Learning-powered Receiver Design for Terahertz Integrated Sensing and Communication Systems}
%\title{A Sensing Integrated DFT-Spread OFDM System for Deep Learning Powered Terahertz Communications}

\author{Yongzhi~Wu,~\IEEEmembership{Graduate~Student~Member,~IEEE,}
        Filip~Lemic,~\IEEEmembership{Member,~IEEE,}
        Chong~Han,~\IEEEmembership{Member,~IEEE,}
        and~Zhi~Chen,~\IEEEmembership{Senior Member,~IEEE}
% <-this % stops a space
%\thanks{This work of Chong Han was supported by the Ministry of Education of China Joint Research Fund under Grant 6141A02033350.
%Filip Lemic’s research stay at the Terahertz Wireless Communications Lab of the Shanghai Jiao Tong University was supported by the Research Foundation - Flanders (FWO), grant nr. V406320N. Filip Lemic was also supported by the EU Marie Curie Actions Individual Fellowship project Scalable Localization-enabled In-body Terahertz Nanonetwork (SCaLeITN), grant nr. 893760.}
% \thanks{This work was presented in part at IEEE Vehicular Technology Conference, 2021~\cite{wu2021dftsofdm}. 
% This work was supported by National Key R\&D Program of China under Grant No. 2020YFB1805700.

% Yongzhi~Wu and Chong~Han are with the Terahertz Wireless Communications (TWC) Laboratory, University of Michigan-Shanghai Jiao Tong University Joint Institute, Shanghai Jiao Tong University, Shanghai 200240, China (Email:~\{yongzhi.wu,~chong.han\}@sjtu.edu.cn).}% <-this % stops a space
% \thanks{Filip Lemic is with the Internet Technology and Data Science Lab (IDLab), University of Antwerpen - imec, Belgium (Email:~filip.lemic@uantwerpen.be).}
% \thanks{Zhi Chen is with University of Electronic Science and Technology of China, Chengdu, China (Email:~chenzhi@uestc.edu.cn).}

\thanks{This work was presented in part at IEEE Vehicular Technology Conference, 2021~\cite{wu2021dftsofdm}.
This work was supported by National Key R\&D Program of China under Project No. 2020YFB1805700. Filip Lemic acknowledges the in-part support by the Spanish Ministry of Economic Affairs and Digital Transformation and the European Union – NextGeneration EU, in the framework of the Recovery Plan, Transformation and Resilience (PRTR) (Call UNICO I+D 5G 2021, ref. number TSI-063000-2021-6-Open6G Joint Open 6G Communications and Sensing).

Yongzhi~Wu and Chong~Han are with the Terahertz Wireless Communications (TWC) Laboratory, Shanghai Jiao Tong University, Shanghai, China (Email:~\{yongzhi.wu,~chong.han\}@sjtu.edu.cn). 

Filip Lemic is with i2Cat Foundation, Spain and University of Antwerpen - imec, Belgium (Email:~filip.lemic@gmail.com).

Zhi Chen is with University of Electronic Science and Technology of China, Chengdu, China (Email:~chenzhi@uestc.edu.cn).}
}

\maketitle
\thispagestyle{empty}


\begin{abstract}
	Terahertz (THz) communications are envisioned as a key technology of next-generation wireless systems due to its ultra-broad bandwidth.
	One step forward, THz integrated sensing and communication (ISAC) system can realize both unprecedented data rates and
	millimeter-level accurate sensing. However, THz ISAC meets stringent challenges on waveform and receiver design to fully exploit the peculiarities of THz channel and transceivers. In this work, a sensing integrated discrete Fourier transform spread orthogonal frequency division multiplexing (SI-DFT-s-OFDM) system is proposed for THz ISAC, which can provide lower peak-to-average power ratio than OFDM and is adaptive to flexible delay spread of the THz channel. Without compromising communication capabilities, the proposed SI-DFT-s-OFDM realizes millimeter-level range estimation and decimeter-per-second-level velocity estimation accuracy. In addition, the bit error rate (BER) performance is improved by 5 dB gain at the 10\textsuperscript{-3} BER level compared with OFDM. At the receiver, a deep learning based ISAC receiver with two neural networks is developed to recover transmitted data and estimate target range and velocity, while mitigating the imperfections and non-linearities of THz systems. Extensive simulation results demonstrate that the proposed deep learning methods can realize mutually enhanced performance for communication and sensing, and is robust against Doppler effects, phase noise and multi-target estimation.
	%under these effects in contrast with the classical methods.
\end{abstract}

\begin{IEEEkeywords}
	Terahertz integrated sensing and communication (THz ISAC), Sensing integrated DFT-spread OFDM (SI-DFT-s-OFDM) waveform, Deep learning (DL).
\end{IEEEkeywords}

\IEEEpeerreviewmaketitle

\section{Introduction}

\IEEEPARstart{I}{n} recent years, exhaustion of spectrum resource in the microwave band has motivated the adoption of higher and wider spectrum. Following this trend of moving up the carrier frequencies, the Terahertz (THz) (0.1-10 THz) is regarded as one of the key technologies for supporting the sixth generation (6G) wireless communication systems. As a highly potential band, 275-450~GHz has been identified by the World Radiocommunication Conference 2019 (WRC-19) for the land mobile and fixed services applications~\cite{WRC2019}. The ultra-broad bandwidth in the THz band enables ultra-fast data rates of up to hundreds of Gbps and even Tbps, and ultra-high sensing accuracy. Due to the short wavelength of THz wave, THz antennas with small sizes are expected to be implemented and support highly portable and wearable devices~\cite{AKYILDIZ201416}. Moreover, the non-ionization of THz radiation ensures that THz devices are safe to human body~\cite{rappaport2019wireless}. 

Meanwhile, along with the trend towards higher frequencies, 6G wireless communication systems envisage integrating communication and sensing to achieve a promising blueprint, i.e., all things are sensing, connected, and intelligent~\cite{zhang20196g}. It is expected that the same system can simultaneously transmit a message and sense the environment by radio signal. Intuitively, the integration of communication and sensing can enhance spectrum efficiency and reduce hardware costs~\cite{zhang2021enabling}. Moreover, when the signal processing modules and the information of the surrounding environment are shared among communication and sensing, their performance can be mutually enhanced. Therefore, by realizing Tbps links and millimeter-level sensing accuracy, the \textit{THz integrated sensing and communication (ISAC)} is envisioned to guarantee high quality-of-experience (QoE) for various services and applications, such as autonomous driving in vehicle networks, wireless virtual reality (VR), and THz Internet-of-Things (Tera-IoT)~\cite{rappaport2019wireless}. In addition, the THz ISAC can provide diverse sensing services, including sensing, localization, imaging and spectrogram~\cite{sarieddeen2020THz}.

Despite the great promise of the THz ISAC, stringent challenges are encountered as a result of the distinctive features of THz wave propagation and devices. First, from the spectrum perspective, as the free-space propagation loss increases quadratically with frequency, it becomes much stronger in the THz band than in the microwave band. In this case, directional antennas are used to provide high gains and compensate for the severe path loss, which reduce the delay spread and increase the coherence bandwidth of the THz channel~\cite{Han2015multi-ray}. Second, the reflection and scattering losses of the THz ray depend on the angle of incidence and usually result in a strong power loss of a non-line-of-sight (NLoS) path, as well as the decrease of the number of the dominant rays with non-negligible power~\cite{wu2020interference}, which can cause a varying delay spread. Third, along with the increase of carrier frequencies, the Doppler shift, which is proportional to the carrier frequency, becomes larger and thus causes stronger Doppler effects. Fourth, from the transceiver perspective, with the increase in the carrier frequency at the wireless communication transceivers, the overall system performance becomes substantially sensitive to radio frequency (RF) analog front-end impairments. In particular, the power amplifier (PA) efficiency of transmitters in the THz band is more sensitive to the peak-to-average power ratio (PAPR) of the transmit signal, since the saturated output power of PA rapidly decreases as the carrier frequency increases~\cite{wang2020pa}. In order to maximize the transmit power and power efficiency, lower PAPR is required to provide higher coverage and promote energy-efficient THz communications. Fourth, there exist phase noise (PN) effects in the local oscillator during the up-conversion and down-conversion of the THz transceivers. Since the PN increases by 6 dB for every doubling of the carrier frequency~\cite{dahlman5g}, it becomes significant to consider the increased PN distortion effect on the THz communications.
THz ISAC systems need to be well designed in terms of the aforementioned challenges, even at the low-THz spectrum. Microwave and millimeter-wave (mmWave) systems are not designed to fully consider all of these challenges and do not work well at the THz spectrum.

\subsection{Related Work}

The concept of integrated sensing and communication has been extensively studied in the literature. Existing papers on integrated sensing and communication can be classified into four classes according to the level of integration~\cite{wu2021ISCI}.
From bottom up, the first level is the communication and sensing coexistence, where the spectrum crunch encourages to share the same frequency bands among communication and sensing~\cite{liu2020jrc}. A typical scenario at this level is a communication system sharing spectrum with a co-located radar system~\cite{zheng2019coexistence,qian2018joint,andrea2020comm}. In this case, the interference is a major issue for the communication and sensing coexistence and thus, efficient interference management techniques are required to avoid the conflict of these two functionalities~\cite{liu2020jrc}. Second, in addition to shared spectrum, when the hardware is shared, the integration of communication and sensing is achieved at a higher level in the dual-functional communication-sensing systems~\cite{zhang2021enabling}. A direct way to implement such a system is to design a time-sharing scheme~\cite{Petrov2019vehicular, zhang2021jcs} or a beam-sharing scheme~\cite{chen2020jsc}, which reduces the cost, size and weight of the system. Alternatively, a common transmitted waveform can be jointly designed and used for communication and sensing, including integrating communication information into radar~\cite{zhang2016radar} and realizing sensing in communication systems~\cite{Kumari2018radar,Strum2011Waveform,Berger2010ofdm,gaudio2020otfs}. The usage of communication waveforms, such as single-carrier~\cite{Kumari2018radar}, orthogonal frequency division multiplexing (OFDM)~\cite{Strum2011Waveform,Berger2010ofdm}, orthogonal time frequency space (OTFS) modulations~\cite{gaudio2020otfs}, are applied to radar sensing and perform as well as the frequency-modulated continuous wave (FMCW) radar in terms of the sensing accuracy. Moreover, future ISAC systems are expected to enable shared signal processing modules at the receiver and further prompt communication and sensing to assist each other~\cite{feng2020jrc}. The fourth level of integration includes the shared protocol and network design above the physical layer.

When moving to higher frequencies, i.e., millimeter wave (mmWave) and THz bands, the directional antenna and beamforming techniques are used to provide high antenna gain and compensate for severe path loss. In this case, communication and sensing have different requirements on beamforming, i.e., sensing requires time-varying directional scanning beams to search the targets in the environment, while by contrast, communication requires accurately-pointed beams to support stable links~\cite{Zhang2019Multi-beam}. Furthermore, the information obtained by sensing can be employed to predict the location of communication receiver in vehicular networks and realize sensing-assisted beamtracking~\cite{liu2020radar}.
In the THz band, a unified framework for vehicular ISAC with a time-domain duplex (TDD) inspired solution is proposed in~\cite{Petrov2019vehicular}, essentially at the second level of integration. Nevertheless, to the best of the authors' knowledge, there are few attempts on higher integration levels of THz ISAC. Motivated by this, our work aims at the third integration level of THz ISAC, by designing a common transmitted ISAC waveform and a deep learning (DL) powered receiver for THz ISAC systems.

As a popular multi-carrier waveform for ISAC in the microwave band, OFDM is well known to be highly spectral-efficient and robust to frequency selective channels~\cite{Zaidi2016waveform} and also has good multiple-input-multiple-output (MIMO) compatibility~\cite{yuan2020hybrid}. Nevertheless, with the increased antenna directivity and reduced delay spread in the THz band, a set of single-carrier waveforms, such as the discrete Fourier transform spread OFDM (DFT-s-OFDM) and its variants~\cite{Sahin2016dft-s-ofdm}, are preferred by the THz systems.
Moreover, low PAPR of the transmit signal is vital for THz transmitters to guarantee effective transmission power and high energy efficiency~\cite{lee2020dftsofdm}.
Thus, DFT-s-OFDM with the single-carrier characteristic is more competitive than OFDM for THz communications.
In our work, we investigate the potential of DFT-s-OFDM for THz ISAC and design a sensing integrated DFT-s-OFDM (SI-DFT-s-OFDM) waveform that is superior to OFDM. Furthermore, we meet two challenges when designing the ISAC receiver. First, there exist strong non-linear distortion effects at the THz transceivers, such as PN effects, which degrade the link performance~\cite{tervo2020thz}, especially when using classical signal recovery methods. Second, it is hard to implement sensing parameter estimation and data detection with one conventional signal processing method.
Nowadays, with a great potential for enhancing performance, deep learning (DL) has been investigated in terms of its applications to communication systems, such as THz indoor localization~\cite{fan2021thz} and channel estimation~\cite{chen2020}. Furthermore, joint channel estimation and signal detection in OFDM systems has been implemented by a deep neural network (DNN) under time-invariant channels, which is more robust to non-ideal conditions than conventional methods~\cite{ye2019dl,gao2018dl}. Existing studies on deep learning for physical layer design work on either the sensing parameter estimation or the communication task separately.
By sharing representations between related tasks, we can enable our DL model to perform better on our original task~\cite{ruder2017mtl}. A deep relationship network with shared convolutional and task-specific fully connected layers is proposed in~\cite{long2017mtl}, which yields superior results on standard datasets in computer vision. In~\cite{lu2017mtl}, an automatic approach that dynamically widens the network and groups similar tasks is developed for person attribute classification. However, the joint parameter estimation approach has not been used in ISAC tasks, since the relation between sensing and communication receivers and the effect brought by the joint parameter estimation model are still not clear. Thus, design of DL-based sensing and communication receivers is still challenging.

\subsection{Contributions}

In light of the aforementioned features of THz channel and transceivers, the THz waveform needs to be well designed to yield a low bit error rate (BER) and a high data rate, as well as to enable accurate sensing capabilities. In this paper, we first propose the SI-DFT-s-OFDM waveform, which maintains the single-carrier characteristic and provides a lower PAPR than OFDM. Furthermore, we address the imperfections of the THz systems, including non-ideal channel conditions and RF impairments, by leveraging the artificial intelligence (AI) techniques, especially deep learning~\cite{Shea2017dl}. To this end, we develop a neural network based ISAC receiver for the THz SI-DFT-s-OFDM system, which can realize mutually enhanced performance for communication and sensing. Remarkably, the proposed waveform and receiver design for THz ISAC are immune to Doppler effect, phase noise and multi-path fading.

The contributions of this work are summarized as follows.
\begin{itemize}
    \item \textbf{We propose a SI-DFT-s-OFDM waveform for the THz ISAC system, by taking into account the peculiarities of the THz channel and transceivers.}
    By designing the frame structure with the data blocks and reference blocks, function of sensing is integrated into this waveform.
    Meanwhile, by considering the varying delay spread of THz channels, we propose a flexible guard interval (FGI) scheme in this waveform, which is able to reduce the cyclic prefix (CP) overhead and improve the data rate.
    The proposed waveform with FGI is able to improve the data rate by tens of Gbps and reduce the PAPR by 3~dB compared to CP OFDM due to its flexibility and single-carrier characteristic.
    The simulation results demonstrate that the proposed SI-DFT-s-OFDM can provide 5 dB gain at the 10\textsuperscript{-3} BER level in the THz channel compared with OFDM.
    
    \item We propose deep learning approaches to sensing parameter estimation and data detection in the THz ISAC systems. Specifically, we first design two preprocessing mechanisms on the received DFT-s-OFDM frames, including block-wise input processing and subcarrier-wise input processing. \textbf{Then we develop a multiple-input average-output sensing neural network (SensingNet) for sensing}, which obtains one prediction result for each input of the received reference signals and implements an average layer to output the final estimation result. The simulation results indicate that the proposed SensingNet offers higher range and velocity estimation accuracy compared with other DL methods. In addition, the developed DL-based sensing receiver can reduce the computational complexity and improve the sensing resolution in contrast with the MUSIC algorithm.
    
    \item \textbf{We develop a two-level communication neural network (ComNet) for data detection.} The first level is designed to extract channel information at the data blocks from the received reference blocks. Then, the received data symbols and the output of the first-level network are concatenated and input into the second-level network to recover transmitted data symbols. In the presence of non-ideal effects including Doppler effects and phase noise, the proposed ComNet achieves better BER performance than existing DL methods and conventional data detection approaches.
    
    \item \textbf{We design a joint sensing and communication neural network by incorporating the proposed SensingNet and ComNet}, which simultaneously estimates sensing parameters and recovers data symbols in the passive sensing. The joint parameter estimation and data detection model can improve the passive sensing accuracy in contrast with the single-task model.
\end{itemize}

The structure of this paper is as follows.
The THz ISAC system model is described in Section~\ref{sec:system_framework}. Section~\ref{sec:waveform} presents the proposed SI-DFT-s-OFDM waveform.
Section~\ref{sec:dnn} delineates the DL-based ISAC receiver. The performance evaluation results are elaborated in Section~\ref{sec:simulation}. Finally, the paper is concluded in Section~\ref{sec:conclusion}.

\section{THz ISAC System Model}\label{sec:system_framework}

In this section, we describe the system model for the proposed sensing integrated DFT-s-OFDM system with the deep learning powered receiver design for THz ISAC. Specifically, we elaborate the system framework, including the transmitter and receiver design, two sensing modes and channel models of THz ISAC systems.

\subsection{System Framework}

\begin{figure*}
    \centering
    \includegraphics[width=0.7\textwidth]{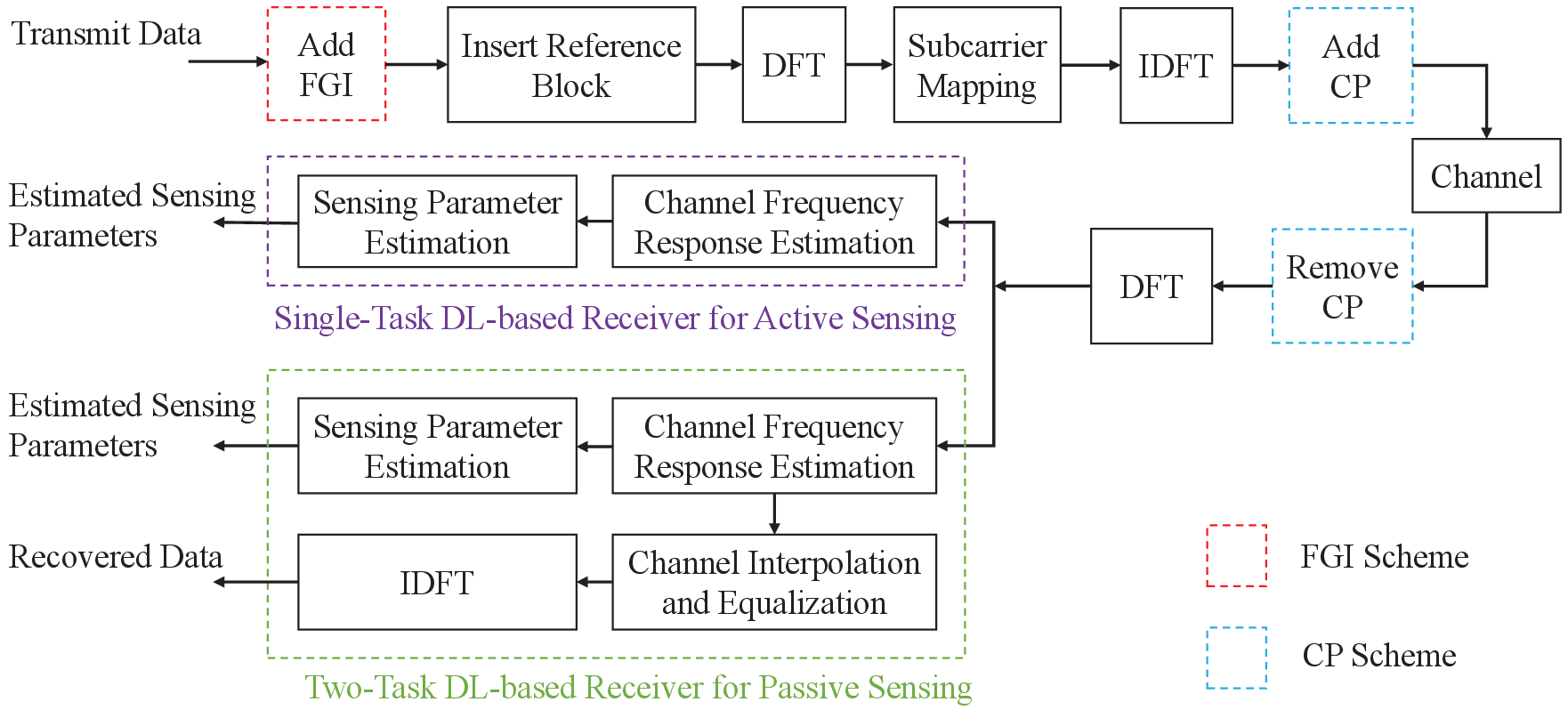}
    \caption{The proposed DFT-s-OFDM system model with deep learning powered receiver design for THz integrated sensing and communication.}
    \label{fig:system_model}
\end{figure*}

As shown in Fig.~\ref{fig:system_model}, we propose a sensing integrated DFT-s-OFDM system with a deep learning powered receiver. At the transmitter (Tx) side, an ISAC waveform is generated and serves for simultaneously enabling communication and sensing functionalities. The ISAC transmitter modulates the transmit data by using the DFT-s-OFDM waveform with two guard interval schemes, i.e., cyclic prefix and flexible guard interval, which will be detailed in Sec.~\ref{sec:waveform}. Meanwhile, the reference blocks are inserted to satisfy the requirements of parameter estimation for sensing and channel estimation for communication. In the time-division scheme~\cite{zhang2021jcs}, the transmit frame consists of sensing subframes and communication subframes, and then the time duration allocation ratio is optimized. While this scheme does not require joint waveform design, the communication channel capacity is sacrificed. Our work aims at designing a shared waveform for sensing and communication, which can improve the efficiency of utilizing time resources and realize higher data rate.

At the receiver side, when using conventional signal processing methods, the channel frequency response (CFR) is estimated by employing the received reference signals. The sensing parameter can be extracted from the estimated channel frequency response by using the radar sensing algorithms~\cite{Strum2011Waveform, Zhang2019Multi-beam}. For data transmission, channel interpolation and equalization can be performed in the frequency domain based on the channel frequency response at the reference blocks. In our work, in order to overcome the aforementioned challenges of THz channel and transceivers, we develop deep learning based sensing and communication receivers to improve the performance of THz ISAC systems.

\subsection{Active and Passive Sensing}

At the receiver (Rx) side, THz ISAC systems can be classified into two sensing modes, i.e., \textit{active sensing} and \textit{passive sensing}.
In the active sensing mode, the transmit signal propagates either through the communication channel to the communication receiver, or through the sensing channel back to the sensing receiver that is collocated with the transmitter. Then the location of the targets can be estimated from the back-reflected return signal. The self-interference from the Tx to the sensing Rx can be suppressed by using the full duplex radar technologies~\cite{barneto2021fullduplex}.
The applications of active sensing includes joint vehicle-to-vehicle communication and radar sensing, and wireless VR communication and sensing.
For active sensing mode, single-task sensing receiver is designed to estimate the range and velocity of targets based on the received reference signals.

The passive sensing system also transmits a signal that is jointly designed and used for communication and sensing. This is followed by the received communication signal serving as the sensing signal that carries the information of the transmitter, such as its distance.
The passive sensing mode of THz ISAC can be applied to some applications, such as THz indoor localization and Tera-IoT, where the communication receivers sense the location of transmitters.
For passive sensing mode, the ISAC receiver are designed to perform joint sensing parameter estimation and data detection by employing the received reference signals and data blocks.

For both sensing modes, we develop a sensing neural network (SensingNet) for sensing receiver and a communication neural network (ComNet) for communication receiver. Furthermore,  for passive sensing, we can incorporate these two networks and design a two-task DL-based ISAC receiver, which estimates sensing parameters and recovers data symbols at the output.

\subsection{Channel Models}\label{sec:channel}

We introduce the channel models for THz ISAC with a \textit{$(N_r+1)$-ray communication channel} model and a \textit{$P$-target sensing channel} model, respectively as follows.
On one hand, the channel impulse response (CIR) of the $(N_r+1)$-ray THz communication channel is \cite{Han2015multi-ray}
	\begin{equation}
	\begin{split}
	h_c(t, \tau) =& \alpha_\text{LoS} e^{j2\pi \nu_{\text{LoS}} t} \delta(\tau - \tau_\text{LoS}) \\
	&+ \sum_{i=1}^{N_r} \alpha_\text{NLoS}^{(i)} e^{j2\pi \nu_\text{NLoS}^{(i)} t} \delta(\tau - \tau_{\text{NLoS}}^{(i)}),
	\end{split}
	\end{equation}
	where $\delta(\cdot)$ denotes the Dirac delta function, $\alpha_\text{LoS}$ and $ \alpha_{\text{NLoS}}^{(i)}$ represent the attenuation for the LoS ray and $i$\textsuperscript{th} NLoS ray, respectively. $N_r$ describes the number of NLoS rays. The propagation delay $\tau_\text{LoS}$ for the LoS ray and $\tau_\text{NLoS}^{(i)}$ for the $i$\textsuperscript{th} NLoS ray can be computed by the equations $\tau_\text{LoS} = \frac{r_\text{LoS}}{c_0}$ and $\tau_\text{NLoS}^{(i)} = \frac{r_\text{NLoS}^{(i)}}{c_0}$, where $r_\text{LoS}$ and $r_\text{NLoS}^{(i)}$ stand for the LoS path distance and the $i$\textsuperscript{th} NLoS path distance, and $c_0$ is the speed of the light. Meanwhile, the time-varying channel response $h_c(t, \tau)$
	is influenced by the Doppler shift $\nu_\text{LoS}$ along the LoS path and $\nu_\text{NLoS}^{(i)}$ along the $i$\textsuperscript{th} NLoS path, which are calculated by $\nu = \frac{f_c v}{c_0}$, where $v$ represents the relative speed between the Tx and the Com Rx along the corresponding path, $f_c$ refers to the carrier frequency.
	
On the other hand, the CIR of the $P$-target sensing channel is described as
	\begin{equation}
	h_\text{s}(t, \tau) = \sum_{p=1}^{P} \alpha_p e^{j2\pi \nu_p t} \delta(\tau - \tau_p),
	\end{equation}
where $P$ is the number of the considered targets, each of which corresponds to one back-reflected path with the attenuation $\alpha_p$. Due to the two-way propagation, the delay and the Doppler shift are calculated by $\tau_p = \frac{2 r_p}{c_0}$ and $\nu_p = \frac{2 f_c v_p}{c_0}$, where $r_p$ and $v_p$ stand for the range and relative speed of the $p$\textsuperscript{th} target, respectively. The speed can be positive or negative, which is determined by the moving direction of the target or the communication receiver. In the radar sensing channel, a negative speed means that a target is moving away. In the communication channel, a negative speed means that the communication receiver is moving away from the transmitter. The power attenuation of communication rays and sensing echoes is calculated as \cite{Han2015multi-ray,richards2014fundamentals}
\begin{subequations}
\begin{align}
|\alpha_\text{LoS}|^2 &= P_t G_{tx} G_{rx} \left(\frac{c_0}{4\pi f_c r_\text{LoS} }\right)^2 e^{-\kappa(f_c) r_\text{LoS} },\\
    |\alpha^{(i)}_\text{NLoS}|^2 &= P_t G^{(i)}_{tx} G^{(i)}_{rx} \left(\frac{c_0}{4\pi f_c r^{(i)}_\text{NLoS} }\right)^2 e^{-\kappa(f_c) r^{(i)}_\text{NLoS} } R_i^2, \\
    |\alpha_p|^2 &= P_t G_{tx}^{(p)} G_{rx}^{(p)} \frac{c_0^2 \sigma_p}{(4\pi)^3 f_c^2 r_p^4} e^{-\kappa(f_c) r_p},
\end{align}
\end{subequations}
where $P_t$ denotes the transmit power, $G_{tx}$ and $G_{rx}$ refer to the transmit and receive antenna gains, the molecular absorption coefficient $\kappa(f_c)$ is a function of the carrier frequency, $R_i$ describes the reflection coefficient and $\sigma_p$ stands for the radar cross section (RCS) of the $p$\textsuperscript{th} sensing target.

In the THz band, directional beams are used to compensate for severe path loss. Sensing prefers scanning beams to search targets in the active sensing, while communication requires stable beams towards the communication receiver~\cite{Zhang2019Multi-beam}. In this case, a fixed sub-beam for communication and several time-varying sub-beams for sensing can be generated by using the THz ultra-massive MIMO (UM-MIMO) and dynamic hybrid beamforming technology~\cite{han2021hybrid}. Multiple target estimation with THz UM-MIMO systems can be extended in future work when considering the additional angle dimension~\cite{elbir2021jrc}.

% In the THz band, multiple antennas should be used to provide high gains and compensate for the severe path loss. We consider the directional transmissions and the antenna gain can be contained in the path gain of the THz channel model. The proposed SI-DFT-s-OFDM waveform is a general solution for THz integrated sensing and communication, regardless of the number of used antennas.

\section{Sensing Integrated DFT-s-OFDM}\label{sec:waveform}

In this section, to reduce the energy efficiency of THz power amplifiers with low saturated power~[10] and integrate high-accuracy sensing into communication, we propose a SI-DFT-s-OFDM waveform with a lower PAPR compared to OFDM.

\subsection{SI-DFT-s-OFDM with Cyclic Prefix}

\begin{figure}[!t]
	\centering
	\subfigure[Tx digital unit of SI-DFT-s-OFDM with CP.]{\includegraphics[width=0.4\textwidth]{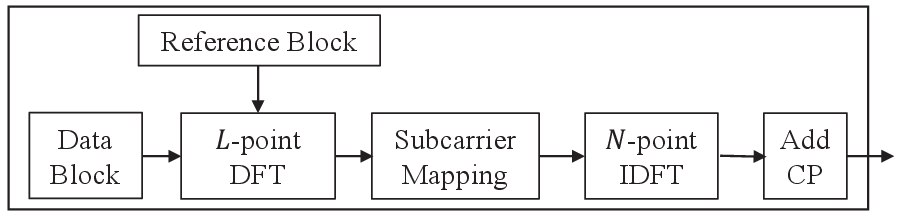}
	}
	\subfigure[Frame design of SI-DFT-s-OFDM with CP.]{\includegraphics[width=0.25\textwidth]{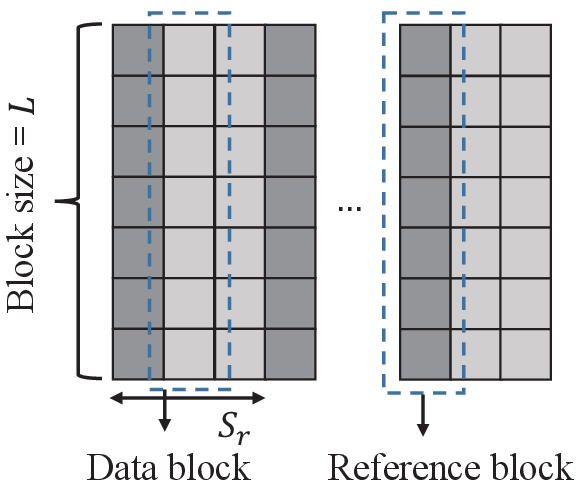}
	}
	\caption{Block diagram of the Tx digital unit and the frame design for the SI-DFT-s-OFDM with CP.}	
	\label{fig:tx_unit_cp}
\end{figure}

As illustrated in Fig.~\ref{fig:tx_unit_cp}, we first introduce the Tx digital unit of the SI-DFT-s-OFDM with CP. At the transmitter, the transmitted data is grouped into multiple data frames. Each data frame with $M$ blocks consists of $M_\text{DB}$ data blocks and $M_\text{RB}$ reference blocks. The input bit streams are firstly mapped to the data sequences with the Q-ary quadrature amplitude modulation (QAM). The modulated symbols are grouped into the data blocks $\mathbf{\tilde{x}}_{\text{D}_m} = [\tilde{x}_{m, 0}, \tilde{x}_{m, 1}, \cdots, \tilde{x}_{m, L - 1}]^T, m = 0, 1, \cdots, M_\text{DB} - 1$, each containing $L$ symbols. Then a $L$-point DFT is performed on $\mathbf{\tilde{x}}_{\text{D}_m}$ and produces a frequency domain representation
\begin{equation}
    \mathbf{\tilde{X}}_{\text{D}_m} = \mathbf{W}_L \mathbf{\tilde{x}}_{\text{D}_m}, m = 0, 1, \cdots, M_\text{DB} - 1,
\end{equation}
where $\mathbf{\tilde{X}}_{\text{D}_m} \triangleq [\tilde{X}_{m, 0}, \tilde{X}_{m, 1}, \cdots, \tilde{X}_{m, L-1}]^T \in \mathbb{C}^{L\times 1}$, and $\mathbf{W}_L \in \mathbb{C}^{L\times L}$ denotes the DFT matrix with the size $L$, $\textbf{W}_L(m, n) \triangleq \frac{1}{\sqrt{L}} \exp\left(-j 2\pi m n / L\right), m, n = 0, 1, \cdots, L - 1$.

The reference blocks are introduced as the sensing and demodulation reference signals, which are generated from constant enveloped Zadoff-Chu (ZC) sequence $\mathbf{p}_{\text{R}_m} = [p_{0}, p_{1}, \cdots, p_{L - 1}]^T, m = 0, 1, \cdots, M_\text{RB} - 1$. Then, the frequency domain representation of the reference block is
\begin{equation}
    \mathbf{P}_{\text{R}_m} = \mathbf{W}_L \mathbf{p}_{\text{R}_m}, m = 0, 1, \cdots, M_\text{RB} - 1,
\end{equation}
which is a sequence with constant envelope. In a data frame of the frequency domain signal, the reference blocks are inserted into the data blocks with equi-distance $S_r$. Thus, the frequency domain SI-DFT-s-OFDM signal is given by
\begin{equation}
    \mathbf{X}_m =
    \begin{cases}
    \mathbf{P}_{\text{R}_q}, m = S_r \cdot q, q = 0, 1, \cdots, M_\text{RB} - 1,\\
    \mathbf{\tilde{X}}_{\text{D}_q}, q = m - q_m, \text{ otherwise},
    \end{cases}
\end{equation}
where $\mathbf{X}_m \triangleq [X_{m, 0}, X_{m, 1}, \cdots, X_{m, L - 1}]^T \in \mathbb{C}^{L \times 1}, m = 0, 1, \cdots, M - 1$, and $q_m$ represents the number of reference blocks before the $m$\textsuperscript{th} block in a frame. The insertion of the reference blocks is used for sensing parameter estimation and signal recovery. Thanks to the ultra-broad bandwidth and ultra-short symbol duration in the THz band, we design the SI-DFT-s-OFDM frame structure with a number of reference blocks, which can achieve high-accuracy sensing.

Next, the subcarrier mapping assigns each block to a set of $L$ consecutive subcarriers and inserts zeros into other $(N - L)$ unused subcarriers. As a result, the time domain SI-DFT-s-OFDM block, $\mathbf{x}_m = [x_{m, 0}, x_{m, 1}, \cdots, x_{m, N - 1}]^T$, is generated by performing an $N$-point IDFT,
\begin{equation}
    \mathbf{x}_m = \mathbf{D}_N \left[\begin{array}{c}
\mathbf{X}_{m} \\
\mathbf{0}_{(N-L) \times 1}
\end{array}\right],
\end{equation}
where $\mathbf{D}_N \in \mathbb{C}^{N\times N}$ refers to the IDFT matrix with size $N$, $\mathbf{D}_N(m, n) = \frac{1}{\sqrt{N}} \exp\left(j2\pi mn/N\right)$, $m, n = 0, 1, \cdots, N - 1$.

In order to avoid the inter-block interference (IBI) caused by the multi-path propagation, a guard interval between adjacent blocks is required. A popular means of dealing with the IBI effect over the multi-path channel is to introduce a cyclic prefix part by copying the last samples of the one block into its front. Let $N_\text{cp}$ denote the length of CP. By adding the CP part, the transmitted blocks become $\mathbf{\hat{x}}_m = [x_{m, N - N_\text{cp}}, \cdots, x_{m, N - 1}, x_{m, 0}, x_{m, 1}, \cdots, x_{m, N-1}]^T$. With the rectangular pulsing shaping and the digital-to-analog conversion, we can further obtain the continuous-time signal as
\begin{equation}
\label{eq_DFT_CP}
    x(t) = \frac{1}{\sqrt{N}}\sum_{m=0}^{M-1} \sum_{n=0}^{L-1} X_{m, n} g(t - m T_o) e^{j2\pi n \Delta f(t - T_\text{cp} - m T_o)},
\end{equation}
where $\Delta f$ represents the subcarrier spacing,  $T_o = T_\text{cp} + T$ refers to the total symbol duration, $T = \frac{1}{\Delta f}$ denotes the original symbol duration, $T_\text{cp} = \frac{N_\text{cp}}{N} T$ stands for the CP duration, $g(t)$ is a rectangular pulse function and equals to 1 for $0 < t < T_o$ and 0 otherwise.

\subsection{SI-DFT-s-OFDM with Flexible Guard Interval}

\begin{figure}[!t]
	\centering
	\subfigure[Tx digital unit of SI-DFT-s-OFDM with FGI.]{\includegraphics[width=0.4\textwidth]{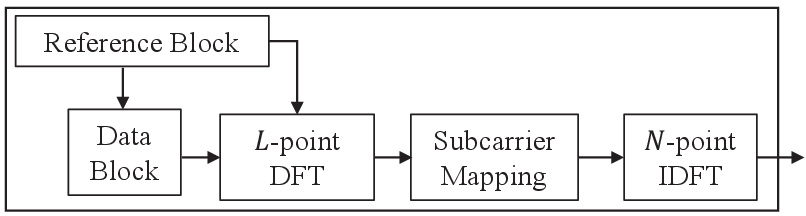}
	}
	\subfigure[Frame design of SI-DFT-s-OFDM with FGI.]{\includegraphics[width=0.35\textwidth]{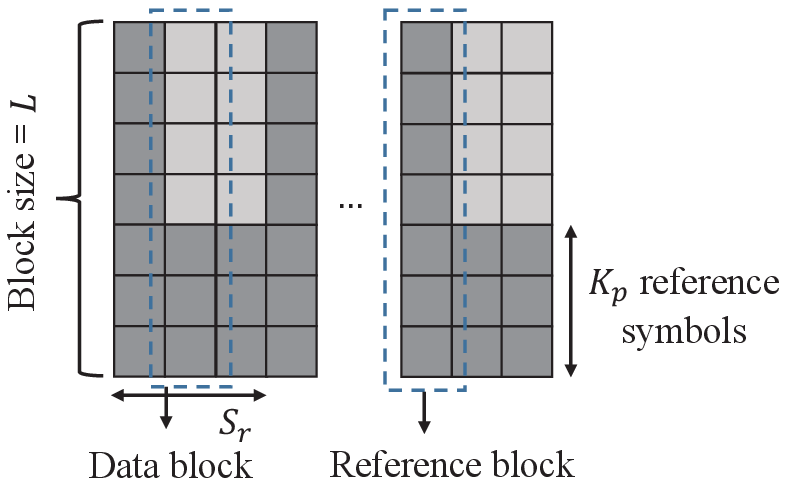}
	}
	\caption{Block diagram of the Tx digital unit and the frame design for the SI-DFT-s-OFDM with FGI.}	
	\label{fig:tx_unit_fgi}
\end{figure}

In current communication systems, the length of CP is usually set longer than the maximum delay spread to remove the IBI effect. However, when it comes to the THz band, the delay spread might fluctuate substantially, e.g., when the signal power of a long NLoS path becomes too weak to influence the received signal, it can be ignored and thereby causes a shorter delay spread. In this case, we can use a short guard interval to reduce the overhead and improve the spectral efficiency.
Nevertheless, the insertion of CP is not flexible, since varying its length may cause different symbol durations and further leads to unfixed frame structure, which makes various settings incompatible.

In order to deal with varying channel delay spread of THz ISAC, we propose the SI-DFT-s-OFDM with FGI, by modifying part of the Tx digital unit. In contrast with copying samples of each data block in the CP scheme, the FGI is generated by the fixed reference symbols and inserted into the data blocks.
As shown in Fig.~\ref{fig:tx_unit_fgi}, when grouping the modulated symbols into the data blocks, we insert a fixed sequence into each data block, which is generated from the tail part of the reference block. In this case, each data block with the block size $L$ is composed of $K$ data symbols and $K_p$ reference symbols, $\mathbf{\tilde{x}}_{\text{D}_m} = [\tilde{x}_{m, 0}, \tilde{x}_{m, 1}, \cdots, \tilde{x}_{m, K - 1}, p_{K}, \cdots, p_{L - 1}]^T$. After performing the $L$-point DFT, subcarrier mapping and $N$-point IDFT, we obtain the time domain block $\mathbf{x}_m = [x_{m, 0}, x_{m, 1}, \cdots, x_{m, N - 1}]^T$ in which the last $K_\text{GI} = \lfloor K_p \frac{ N}{L} \rfloor$ samples are approximately constant, i.e., $x_{i, n} \approx x_{j, n}, n = N - K_\text{GI}, \cdots, N - 1$ for $i\neq j$. 

Based on this feature, we can regard the last $K_\text{GI}$ samples of the $m$\textsuperscript{th} block as the approximate cyclic prefix of the $m$\textsuperscript{th} block, which is essentially an internal guard interval inside the IDFT output. Therefore, we do not need extra operation of adding CP. Meanwhile, by flexibly adjusting the number of data symbols and the length of the fixed sequence with fixed block size, it can satisfy different requirements of guard interval length for the channel delay spread. Without the CP part, the continuous time signal of the SI-DFT-s-OFDM is expressed as
\begin{equation}
\label{eq_DFT_FGI}
      x(t) = \frac{1}{\sqrt{N}}\sum_{m=0}^{M-1} \sum_{n=0}^{L-1} X_{m, n} g(t - m T) e^{j2\pi n \Delta f(t - m T)},
\end{equation}
where the CP part in \eqref{eq_DFT_CP} is replaced by the FGI part $x(t) (mT - \frac{K_\text{GI}}{N} T < t < m T , m = 1, \cdots, M)$ in \eqref{eq_DFT_FGI}.

\section{Deep Learning Based Receiver for THz ISAC}\label{sec:dnn}

\subsection{Signal Pre-processing}

Before developing the deep learning method, we perform pre-processing on the received signal.
Since the channel models for active and passive sensing have similar forms, we can conduct similar analysis on each received block by using a unified baseband channel impulse response $h(t, \tau) = \sum_{l=0}^{N_P-1} h_l e^{j2\pi \nu_l t} \delta(\tau - \tau_l)$, where $N_P$ denotes the number of transmission paths or targets, $h_l$, $\tau_l$ and $\nu_l$ represent the normalized complex path gain, the path delay and the Doppler shift of the $l$\textsuperscript{th} path, respectively.

We derive the received block of the SI-DFT-s-OFDM with CP as follows. The noiseless received signal $r(t)$ through the communication or sensing channel is given by
\begin{equation}
\begin{split}
     r(t) = \int h(t, \tau) x(t - \tau) d\tau = \sum_{l=0}^{N_P - 1} h_l e^{j2\pi \nu_l t} x(t - \tau_l).
\end{split}
\end{equation}
The received noiseless samples of the $m$\textsuperscript{th} block are expressed as $\mathbf{r}_{m} = [r_{m, 0}, r_{m, 1}, \cdots, r_{m, N - 1}]^T$, where
\begin{equation}
\begin{split}
    r_{m, i} &= r(t)|_{t=m T_o + T_\text{cp} + i \frac{T}{N}} \\
    &= \frac{1}{\sqrt{N}}\sum_{l=0}^{N_P - 1} \alpha_l e^{j 2\pi \nu_l mT_o} \sum_{n=0}^{L-1} X_{m, n} e^{j2\pi n \Delta f (i\frac{T}{N} - \tau_l)},
\end{split}
\end{equation}
where $\alpha_l = h_l e^{j2\pi \nu_l \left(T_\text{cp} + i \frac{T}{N} \right)} \approx h_l e^{j2\pi \nu_l T_\text{cp}}$.
In the presence of phase noise and additive white Gaussian noise (AWGN), the noisy received block $\mathbf{y}_m = [y_{m, 0}, y_{m, 1}, \cdots, y_{m, N-1}]^T$ is given by
\begin{equation}
    \mathbf{y}_m = \mathbf{Q}_m \mathbf{r}_m + \mathbf{z}_m,
\end{equation}
where $\mathbf{Q}_m \in \mathbb{C}^{N\times N}$ refers to the phase noise effect in the THz band and is a diagonal matrix given by
$
    \mathbf{Q}_m = \text{diag}\{[e^{j\theta_{m, 0}}, e^{j\theta_{m, 1}}, \cdots, e^{j\theta_{m, N - 1}}]\}
$,
where $e^{j\theta_{m, n - 1}}$ represents the phase noise at the $n$\textsuperscript{th} samples of $m$\textsuperscript{th} received block. Besides, $\mathbf{z}_m$ denotes the $m$\textsuperscript{th} AWGN vector.

The phase noise process is modeled as the Wiener process~\cite{mehrpouyan2012pn}, which is given by
\begin{equation}
    \theta_{m, n} = \theta_{m, n-1} + \Delta \theta_{m, n},
\end{equation}
where $\Delta \theta_{m, n}$ follows the real Gaussian distribution, $\mathcal{N}(0, \sigma^2_\theta)$. The variance $\sigma^2_\theta$ is calculated by $\sigma^2_\theta = 2\pi f_\text{3dB} T_s$, where $f_\text{3dB}$ denotes the one-sided 3-dB bandwidth of the Lorentzian spectrum of the oscillator at the receiver and $T_s$ represents the sampling duration.
With the increase of the carrier frequencies, phase noise effects in the local oscillator become stronger in the THz band.

When the CP part is replaced by the flexible guard interval, we derive the received samples of the $m$\textsuperscript{th} block as
\begin{equation}\label{eq:rxsignal}
    \begin{split}
    r_{m, i} =& r(t)|_{t=m T + i \frac{T}{N}} \\
    =& \sum_{l=0}^{N_P - 1} h_l e^{j2\pi \nu_l t} x(t - \tau_l)|_{t=m T + i \frac{T}{N}} \\
    % =& \frac{1}{\sqrt{N}}\sum_{l=0}^{N_P - 1} h_l e^{j2\pi \nu_l (m T + i \frac{T}{N})} \left(\sum_{n=0}^{L-1} X_{m, n} g\left(\frac{i}{N} T - \tau_l\right)e^{j2\pi n\Delta f(\frac{i}{N} T - \tau_l)} \right.\\
    % &\left.+ \sum_{n=0}^{L-1} X_{m-1, n} g\left(T + \frac{i}{N} T - \tau_l\right)e^{j2\pi n\Delta f(T + \frac{i}{N} T - \tau_l)}\right) \\
    =& \frac{1}{\sqrt{N}}\sum_{l=0}^{N_P - 1} h_l e^{j2\pi \nu_l (m T + i \frac{T}{N})} \sum_{n=0}^{L-1} \left(X_{m, n}\mathbb{I}\left(\tau_l \leqslant \frac{i}{N}T \right) \right.\\
    &\left.+ X_{m -1, n} \mathbb{I}\left(\tau_l > \frac{i}{N} T\right)\right) e^{j2\pi n \Delta f (\frac{i}{N} T - \tau_l)},
    \end{split}
\end{equation}
where $\mathbb{I}\left(\cdot\right)$ refers to the indicator function.
We observe that the SI-DFT-s-OFDM with FGI does not use the perfect cyclic prefix, which may cause weak IBI due to the propagation paths with long delay.

In order to conduct the sensing parameter estimation and the data detection, we need to perform $N$-point DFT operation on $\mathbf{y}_m$ and subcarrier demapping. As a result, the received frequency domain signal is written as
\begin{equation}\label{eq:rx_freq_signal}
    \mathbf{Y}_m = \left[\begin{array}{cc}
\mathbf{I}_{L} &
\mathbf{0}_{L \times (N-L)}
\end{array}\right] \mathbf{W}_N \mathbf{y}_m,
\end{equation}
where $\mathbf{Y}_m \triangleq [Y_{m, 0}, Y_{m, 1}, \cdots, Y_{m, L - 1}]^T$. Furthermore, we deduce the frequency domain representation of the received reference block and the received data block, respectively as $\mathbf{\tilde{P}}_{\text{R}_m} \triangleq [\tilde{P}_{m, 0}, \tilde{P}_{m, 1}, \cdots, \tilde{P}_{m, L - 1}]^T $ and $\mathbf{\tilde{Y}}_{\text{D}_m} \triangleq [\tilde{Y}_{m, 0}, \tilde{Y}_{m, 1}, \cdots, \tilde{Y}_{m, L - 1}]^T$.

\subsection{Analysis of Sensing and Communication Tasks}

In the SI-DFT-s-OFDM system, the reference blocks have a constant envelop in both time and frequency domains, which can be used for the aforementioned channel estimation based sensing algorithms. Meanwhile, they are usually generated by a fixed sequence, which is assumed to be known by both transmitters and receivers. Thanks to the very short symbol duration of THz waveform, a number of reference blocks can be inserted into a data frame, which contributes to high sensing accuracy.
The received frequency domain reference signals at the sensing receiver are given by
$
\tilde{P}_{m, n}^{(s)} =  H_{m, n}^{(s)} P_{m, n} + Z_{m, n}^{(s)},
$
where $m = 0, 1, \cdots, M_\text{RB} - 1 \text{, and } n = 0, 1, \cdots, L - 1$, $Z_{m, n}^{(s)}$ refers to the AWGN, and $H_{m, n}^{(s)}$ denotes the sensing CFR at the $n$\textsuperscript{th} subcarrier of $m$\textsuperscript{th} reference block, derived as
$
    H_{m, n}^{(s)} \overset{}{\approx} \sum_{p = 1}^P \alpha_p e^{j2\pi \nu_p m S_r T_o } e^{-j2\pi \tau_p n \Delta f}.
$
Next, we can perform the LS channel estimation and obtain the estimated CFR, which is expressed as
\begin{equation}
    \hat{H}^{(s)}_{m, n} = \frac{\tilde{P}^{(s)}_{m, n}}{P_{m, n}} = H^{(s)}_{m, n} + \frac{Z^{(s)}_{m, n}}{P_{m, n}}.
\end{equation}
The sensing CFR at the reference blocks is then regarded as the sensing processing matrix. Due to the constant envelop of $P_{m, n}$, the LS estimator does not increase the noise variance.

Then, the sensing task is to estimate the delay and Doppler parameters, $\tau_p$ and $\nu_p$, from estimated sensing CFR, and calculate the range and velocity parameters. This estimation problem can be expressed as a problem of spectral estimation from a sum of complex exponential signals buried in noise.
Furthermore, both the range and velocity can be estimated from the correlation function by using the high-resolution subspace-based methods, such as multiple signal classification (MUSIC)~\cite{le2017ofdm}. Alternatively, the DFT-based method~\cite{Strum2011Waveform} can be invoked as the sensing algorithm, which is the maximum likelihood estimator and performs DFT on the sensing CFR.
The Cram\'{e}r-Rao lower bounds (CRLBs) for range and velocity estimation variance in case of one target using frequency domain signals can be seen in~\cite{braun2014ofdm}.

At the communication receiver, several steps are implemented to perform the communication task, including IDFT/DFT operations, channel estimation and channel equalization. In the ISAC system, the reference blocks are not only used for the sensing parameter estimation, but also for the frequency domain equalization (FDE) at the communication receiver, including the zero-forcing (ZF) and the minimum mean square error (MMSE) equalization methods.

The disadvantages of conventional signal processing methods include limited robustness to non-linear distortions, e.g., Doppler effect and PN noise, and difficulty to simultaneously perform sensing and communication, causing that an integrated receiver for ISAC is challenging.
Thus, we delineate the DL-based ISAC receiver to estimate the sensing parameters and recover the communication data in THz SI-DFT-s-OFDM systems, which can overcome the above problems.

\begin{figure*}
    \centering
    \subfigure[Range parameter estimation.]{\includegraphics[width=0.6\textwidth]{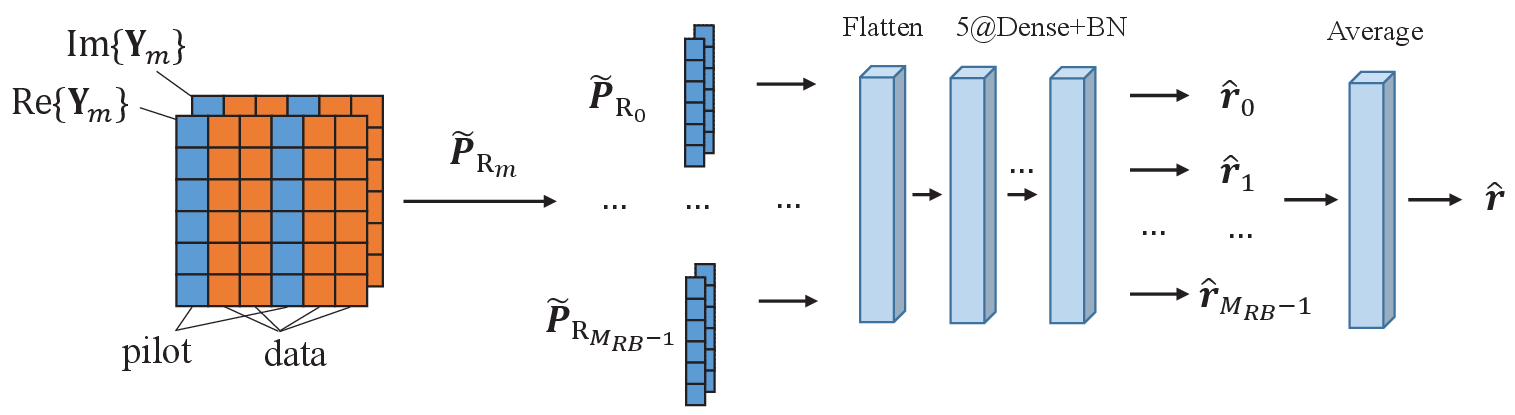}\label{fig:SensingNet_range}}
    \subfigure[Velocity parameter estimation.]{\includegraphics[width=0.6\textwidth]{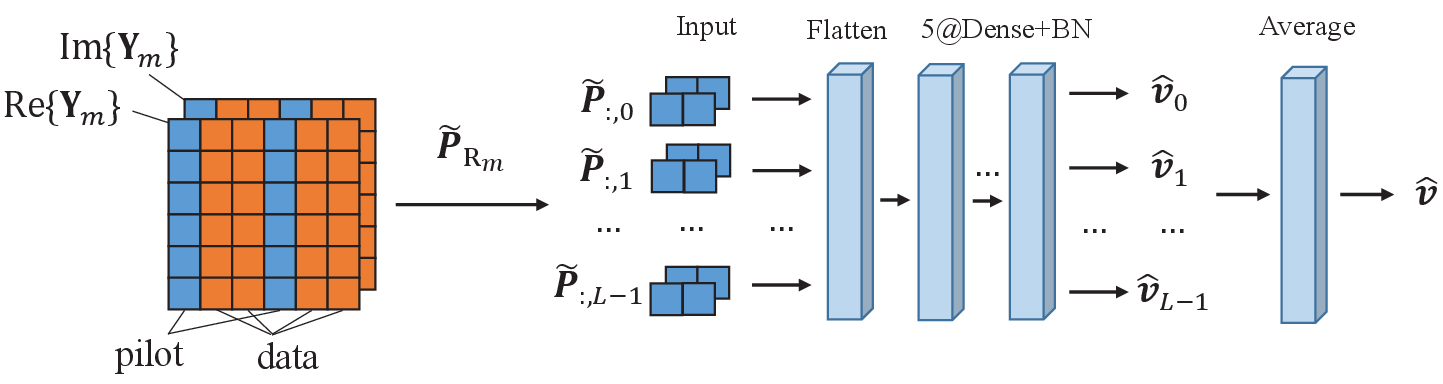}\label{fig:SensingNet_velocity}}
    \caption{The structure of the proposed SensingNet network for DL-based sensing receiver.}
    \label{fig:SensingNet}
\end{figure*}

\subsection{Sensing Neural Network for Sensing Parameter Estimation}

To estimate the sensing parameters, a multiple-input average-output SensingNet network shown in Fig.~\ref{fig:SensingNet} is developed in this section, which regards the received reference signals as input features. In the designed SensingNet network, we adopt block-wise input processing for range estimation and subcarrier-wise input processing for velocity estimation. Thus, the SensingNet models have $M_\text{RB}$ inputs for range estimation and $L$ inputs for velocity estimation. Dense layers are deployed and to obtain one prediction result for each input, which shares the same dense layers. After that, an average layer is implemented to complete the SensingNet network and output the final estimation result.

\subsubsection{Range Estimation}
In Fig.~\ref{fig:SensingNet_range}, the SensingNet network for range estimation consists of an input layer, a flatting layer, five dense layers for feature extraction and nonlinear mapping, and an average output layer. The input layer has $M_\text{RB}$ inputs and each input is composed of two real-valued vectors, which corresponds to one block of the received reference signals $\tilde{\mathbf{P}}_{\text{R}_m}$, i.e., the element-wise real and imaginary values denoted by $\text{Re}\{\tilde{\mathbf{P}}_{\text{R}_m}\}$ and $\text{Im}\{\tilde{\mathbf{P}}_{\text{R}_m}\}$.  Following the input layer, a flatting layer rearranges each input into one dimension and connects to five dense layers, which extract the delay information of the THz sensing channel from each input. The number of neurons in each dense layer used for SensingNet are 500, 250, 120, 60, $P$. In addition, the batch-normalization (BN) operation is invoked at each dense layer to prevent overfitting. Therein, the last dense layer exports an estimated range vector with $P$ range parameters for each input, $\hat{\mathbf{r}}^{(m)}, m = 0, 1, \cdots, M_\text{RB} - 1$. Finally, the average output layer calculates an average of the $M_\text{RB}$ range vectors, namely, the element-wise mean value, $\hat{\mathbf{r}} = \frac{1}{M_\text{RB} } \sum_{m=0}^{M_\text{RB}-1} \hat{\mathbf{r}}^{(m)}$. By denoting the activation function of the $i$\textsuperscript{th} dense layer as $f^{(i)}$, the network output revealing the estimated range vector can be represented as
\begin{equation}
    \hat{\mathbf{r}} = \frac{1}{M_\text{RB} } \sum_{m=0}^{M_\text{RB}-1} f^{(5)} \left(\cdots f^{(1)}\left(\left[\text{Re}\{\tilde{\mathbf{P}}_{\text{R}_m}\}; \text{Im}\{\tilde{\mathbf{P}}_{\text{R}_m}\}\right]\right)\right).
\end{equation}
The rectified linear unit (ReLU) activation function, $
    f_\text{ReLU}(x) = \max (0, x)
$, is introduced to implement non-linear mapping and speed up the computation at the first 4 dense layers.
The sigmoid activation function, $
    f_\text{Sigmoid}(x) = \frac{1}{1 + e^{-x}}
$,
which maps the output to the interval $[0, 1]$, is employed at the last dense layer. During the network training, we leverage supervised learning to minimize the mean square error (MSE) loss function as
\begin{equation}
    \text{Loss}_r = \frac{1}{N_\text{train}} \sum_{\mathbf{r}\in \mathcal{D}} \left\|\mathbf{r} - \hat{\mathbf{r}}\right\|_2^2,
\end{equation}
where $\mathcal{D}$ denotes the network training dataset that contains $N_\text{train}$ samples and the label vector is composed of the true range parameters, $
    \mathbf{r} = [r_1, r_2, \cdots, r_P]^T
$.

\begin{figure*}
    \centering
    \includegraphics[width=0.7\textwidth]{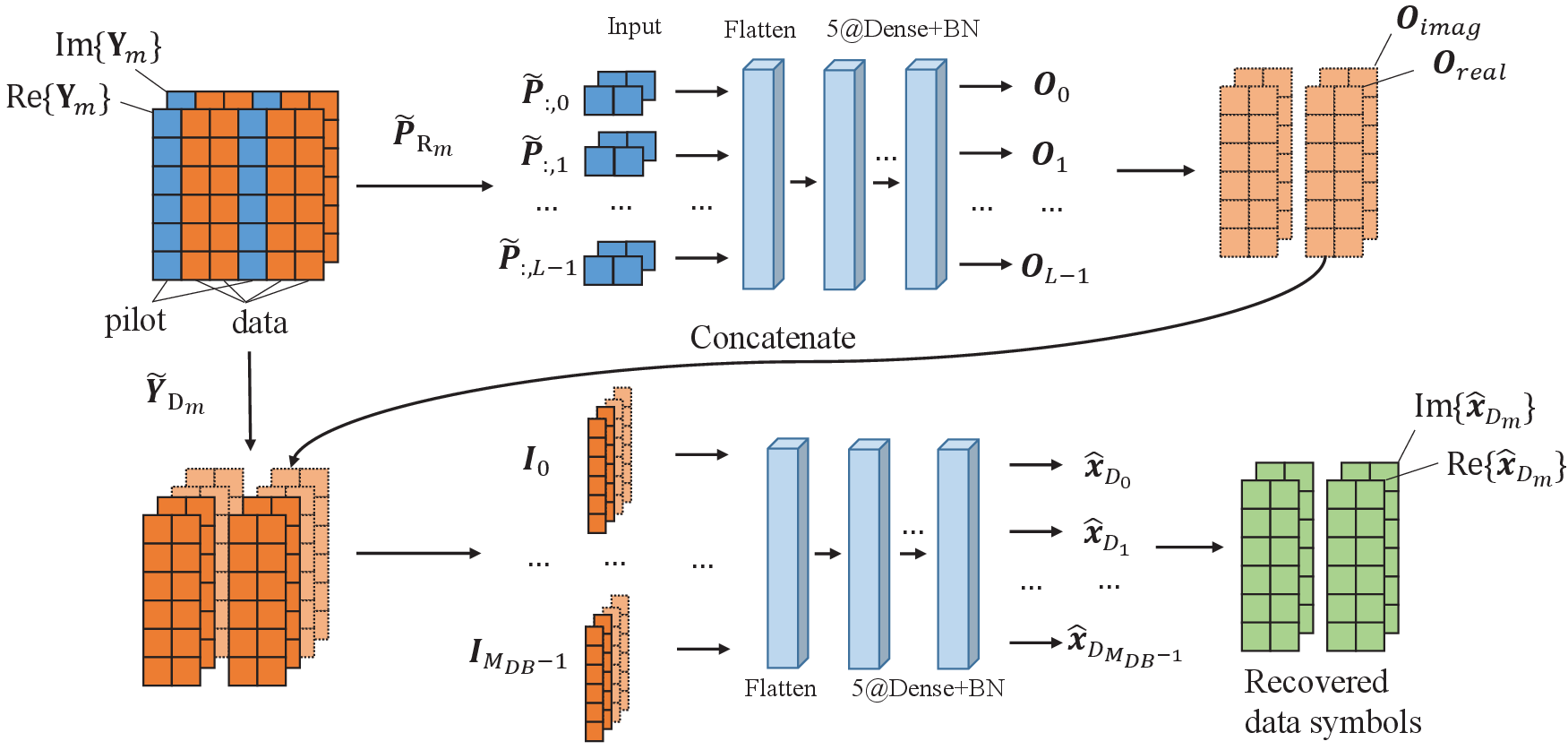}
    \caption{The structure of the proposed two-level ComNet network for DL-based communication receiver.}
    \label{fig:ComNet}
\end{figure*}

\subsubsection{Velocity Estimation}
As shown in Fig.~\ref{fig:SensingNet_velocity}, the SensingNet model for velocity estimation has similar network structure as the model for range estimation and conduct subcarrier-wise input processing at the input layer. The input layer has $L$ inputs, each of which is formed by the received reference signals at a subcarrier, $\tilde{\mathbf{P}}_{:, n} =[\tilde{P}_{0, n}, \tilde{P}_{1, n}, \cdots, \tilde{P}_{M_\text{RB} - 1, n}]^T, n = 0, 1, \cdots, L - 1$. In the same way, the last dense layer deduces an estimated velocity vector for each input and the average layer outputs a mean velocity vector. Therein, the network output predicting the velocity vector can be expressed as
\begin{equation}
    \hat{\mathbf{v}} = \frac{1}{L } \sum_{n=0}^{L-1} f^{(5)} \left(f^{(4)}\left(\cdots f^{(1)}\left(\left[\text{Re}\{\tilde{\mathbf{P}}_{:, n}\}; \text{Im}\{\tilde{\mathbf{P}}_{:, n}\}\right]\right)\right)\right).
\end{equation}
The hyperbolic tangent function, $
    f_\text{tanh} = \frac{e^{x} - e^{-x}}{e^{x} + e^{-x}}
$, is employed as the activation function at the last dense layer of the SensingNet for velocity estimation. To train the velocity estimation network, the loss function $\text{Loss}_v$ becomes the MSE between the estimated result $\hat{\mathbf{v}}$ and the true velocity vector, $\mathbf{v} = [v_1, v_2, \cdots, v_P]^T$.

When training the SensingNet network, we perform the normalization operation on the sensing label. Specifically, we map the target range value into the interval [0, 1], and the target velocity into the interval [-1, 1].
The optimizer used for training our network is the adaptive moment estimation (Adam)~\cite{kingma2014adam}, which is a combination of root mean square propagation (RMSprop) and stochastic gradient descent (SGD) with momentum, due to its fast convergence speed and higher computational efficiency compared to other SGD methods~\cite{kingma2014adam}.
In practical applications, the channel parameters are randomly generated within some range. The model needs to be re-trained when the parameters are out of this range. Nevertheless, since all training processes can be conducted at the offline stage, the feasibility of DL-based methods would not be influenced.

\subsection{Communication Neural Network for Data Detection}

As a deep learning-based communication receiver, a two-level ComNet network shown in Fig.~\ref{fig:ComNet} is proposed to recover the data symbols, which is able to improve the robustness to Doppler effects and phase noise. In the developed ComNet network, the first level is designed to extract channel information at the data blocks from the received reference blocks. Then the received data symbols and the output of the first-level network are concatenated and input into the second-level network. After that, the second level network is deployed to approximate the relationship between the received data symbols and the transmitted data symbols and output the recovered data.

\subsubsection{Network Structure of the First Level}
In Fig.~\ref{fig:ComNet}, the first level sub-network of the ComNet consists of an input layer and five dense layers. Performing like the velocity estimation SensingNet model, the input layer conduct subcarrier-wise input processing on the received reference signal, where each input fully describes the channel observation at each subcarrier. Following the input layer, five dense layers are employed to extract the channel Doppler features from the reference blocks and export the channel frequency response information at the data blocks. The layer sizes of each dense layer are respectively 500, 250, 120, 60, $2 M_\text{DB}$. Therein, the last dense layer outputs $2 M_\text{DB}$ values for each input, which corresponds to the real and imaginary values that contains information of CFR at the data signals. By denoting the activation function of the $i$\textsuperscript{th} dense layer as $g^{(i)}$, the $n$\textsuperscript{th} output of the first sub-network can be represented as
\begin{equation}
    \mathbf{O}_n = g^{(5)} \left(g^{(4)}\left(\cdots g^{(1)}\left(\left[\text{Re}\{\tilde{\mathbf{P}}_{:, n}\}; \text{Im}\{\tilde{\mathbf{P}}_{:, n}\}\right]\right)\right)\right),
\end{equation}
where $\mathbf{O}_n \in \mathbb{R}^{2 M_\text{DB} \times 1}, n = 0, 1, \cdots, L - 1$. In the first sub-network, the activation functions are $f_\text{tanh}$ for the last dense layer and $f_\text{ReLU}$ for other dense layers. Then, $L$ outputs from the the first sub-network are reshaped and concatenated into two matrices, denoted by
\begin{align}
    \mathbf{O}_\text{real} = [\mathbf{O}_0', \mathbf{O}_1', \cdots, \mathbf{O}_{L-1}' ]^T, \\
    \mathbf{O}_\text{imag} = [\mathbf{O}_0'', \mathbf{O}_1'', \cdots, \mathbf{O}_{L-1}'' ]^T,
\end{align}
where $\mathbf{O}_n' \in \mathbb{R}^{M_\text{DB}\times 1}$ and $\mathbf{O}_n'' \in \mathbb{R}^{M_\text{DB}\times 1}$ stand for the first half part and the last half part of $\mathbf{O}_n$, respectively.

\subsubsection{Network Structure of the Second Level}
In the designed ComNet network, the output of the first level and the received data symbols are concatenated and then input into the second level sub-network. The block-wise input processing is performed at the second sub-network, i.e., the $m$\textsuperscript{th} input can be expressed as
\begin{equation}
    \mathbf{I}_m = \left[\text{Re}\{\tilde{\mathbf{Y}}_{\text{D}_m}\}; \text{Im}\{\tilde{\mathbf{Y}}_{\text{D}_m}\}; \mathbf{O}_\text{real}^{(m)}; \mathbf{O}_\text{imag}^{(m)}\right],
\end{equation}
where $\mathbf{O}_\text{real}^{(m)}$ and $\mathbf{O}_\text{imag}^{(m)}$ refer to the $m$\textsuperscript{th} column vector of $\mathbf{O}_\text{real}$ and $\mathbf{O}_\text{imag}$, respectively, $m = 0, 1, \cdots, M_\text{DB} - 1$. By deploying dense layers in the second level sub-network, the $m$\textsuperscript{th} output is denoted by
\begin{equation}
    \left[\text{Re}\{\hat{\mathbf{x}}_{\text{D}_m}\}; \text{Im}\{\hat{\mathbf{x}}_{\text{D}_m}\}\right] = g^{(10)} \left(g^{(9)}\left(\cdots g^{(6)}\left(\mathbf{I}_m\right)\right)\right),
\end{equation}
where $f_\text{tanh}$ and $f_\text{ReLU}$ are respectively used for the last dense layer and other dense layers. Thus, the final output of the ComNet network is composed of $M_\text{DB}$ recovered data vectors. The proposed ComNet network can be also applied to OFDM waveform, since DFT-s-OFDM is compatible with OFDM system.

The loss function of training the ComNet network is defined as
\begin{equation}
    \text{Loss}_c = \frac{1}{N_\text{train} M_\text{DB}} \sum_{\tilde{\mathbf{x}}_{\text{D}_m}\in \mathcal{D}} \left\|\tilde{\mathbf{x}}_{\text{D}_m} - \hat{\mathbf{x}}_{\text{D}_m}\right\|_2^2,
\end{equation}
where $\mathbf{\tilde{x}}_{\text{D}_m}$ and $\mathbf{\hat{x}}_{\text{D}_m}$ refer to the true and the estimated label vectors for data detection. When training the ComNet network models, for each data frame that contains $L \times M_\text{DB}$ data symbols, we divide the whole data symbols into $\frac{1}{8} L$ groups. Every $8 M_\text{DB}$ transmitted data symbols are grouped and predicted based on a single model trained independently, which is then concatenated for the final output data frame. This mechanism is commonly used in the DL-based communication receivers~\cite{ye2019dl, gao2018dl}.
Moreover, different from some work that classifies the transmitted bits~\cite{ye2019dl}, we use regression algorithms to recover the QAM data symbols. In this case, the size of the outcome of the learning model would not be very large the convergence process is not dramatically time-consuming. Specifically, the loss function of sensing reduces by three orders of magnitude after 40 epochs and the loss function of communication decreases by two orders of magnitude after 80 epochs, which verify the fast convergence of the proposed deep learning methods.

In the THz ISAC systems, we have developed two neural network methods for DL-based ISAC receiver, namely, SensingNet and ComNet. In the passive sensing mode, the ISAC receiver is required to simultaneously estimate sensing parameters and recover data symbols, which can be implementated by directly using separate SensingNet and ComNet network models. Alternatively, a joint sensing and communication neural network (ISACNet) can be designed by incorporating SensingNet and ComNet. Since the ComNet and SensingNet has similar input forms and network structure, the two-task neural network can be implemented by using shared layers and non-shared layers. Therein, the hard parameter sharing is employed in the shared layers to reduce the network parameters and possibility of overfitting. While the part of shared layers shares significant knowledge about wireless channel between communication and sensing, the non-shared layers contains two task-specific sub-networks that optimize communication and sensing performance, respectively. 
To train the multi-task network model, we can use the weighted sum method to define the loss function, denoted by
\begin{equation}
    \text{Loss} = a_1 \text{Loss}_c + a_2 \text{Loss}_r + a_3 \text{Loss}_v,
\end{equation}
where $a_1, a_2, a_3$ stand for the weights of the loss functions for data detection, range estimation and velocity estimation, respectively.

While tasks are related, the quantification of the task relatedness is still an open issue. In most applications, task relations are not available or assumed to be known as a priori information~\cite{zhang2021mtl}.
In our work, the measure of task relatedness between the hypothesis of sensing and the hypothesis of communication is defined as~\cite{daniel1996relatedness},
\begin{equation}\label{eq:relatedness}
    R = \text{tanh}\left(\frac{\eta a}{d^2 + \epsilon}\right),
\end{equation}
where the tanh function restricts the value of $R$ to the range (0, 1), $a = \frac{1}{\text{Loss}}$ quantifies the accuracy of two tasks, $d$ refers to the weight space distance between sensing and communication layers, $\epsilon$ is a small constant to prevent division by 0, and $\eta$ controls the rate of decay of tanh from 1.

\section{Simulation Results}\label{sec:simulation}

\begin{table}
    \centering
    \caption{Simulation Parameters}
    \label{tab:parameters}
    \begin{tabular}{ccc}
    \toprule
        \textbf{Notation} & \textbf{Definition} & \textbf{Value}  \\
        \midrule
        $f_c$ & Carrier frequency & 0.3 THz \\
        $\Delta f$ & Subcarrier spacing & 1.92, 7.68 MHz \\
        $T$ & Symbol duration & 0.13, 0.52 $\mu s$ \\
        $T_\text{cp}$ & CP duration & 0.032, 0.13 $\mu s$ \\
        $N$ & Subcarrier number & 64, 256, 1024 \\
        $L$ & Block size & 32, 128, 512 \\
        $\mathbb{A}$ & Modulation scheme & 4-QAM \\
        $\sigma_\theta^2$ & PN variance & 10\textsuperscript{-4}, 10\textsuperscript{-3}, 10\textsuperscript{-2} \\
    \bottomrule
    \end{tabular}
\end{table}

In this section, we investigate the performance of the proposed THz ISAC system with the SI-DFT-s-OFDM waveform and the DL-powered receiver, in contrast with OFDM and conventional signal processing methods. The key parameters in simulations are described in Table~\ref{tab:parameters}. The subcarrier spacing is set as $15\times 2^n \text{ kHz}$ to be compatible with 4G and 5G numerology~\cite{Zaidi2016waveform}. In addition, we refer to the THz link budget analysis in~\cite{rikkinen2020thz} for other parameters.

\subsection{Generation of the Dataset}

The training dataset is generated by using the channel models introduced in Sec.~\ref{sec:channel}in the simulated environment. In particular, we first set system parameters, including waveform parameters given in Table~\ref{tab:parameters} and maximum values of channel parameters. Next, for each data sample, one transmit frame with random data is generated according to the proposed waveform scheme in Sec.~\ref{sec:waveform}. Then, we construct a channel impulse response by generating several propagation paths with delays that are randomly selected between zero and guard interval duration. Moreover, the speed is randomly distributed within the range of [-100, 100] km/h. In this case, a point-to-point communication simulation can be conducted by calculating the channel output. Meanwhile, the AWGN and the phase noise are considered to evaluate the robustness of the proposed DL methods to these effects. Finally, we can obtain the received signals and regard them as the input features of dataset. The transmitted data symbols and the channel parameters can be viewed as the labels of dataset. The SNR value of training dataset is fixed as 25 dB. In addition, the size of training dataset is 50k, while the size of test dataset is 5k.
All the DL-based solutions are implemented in Python and datasets are generated in the Matlab environment. The software and hardware settings of the experiments are provided in Table~\ref{tab:software}.

\begin{table}[t]
    \centering
    \caption{Software and hardware setting for the experiments.}\label{tab:software}
    \begin{tabular}{cc}\toprule
        \textbf{Item} & \textbf{Version} \\
        \midrule
        GPU & 1 NVIDIA GEFORCE RTX 2080 Ti \\
        CPU & 9th Generation Intel Core i7 Processors \\
        Python version & Python 3.6\\
        Matlab version & Matlab R2018a \\
        Deep learning framework & Tensorflow 1.14.0 \\
        CUDA/cuDNN version & CUDA 10.0 and cuDNN 7.4.1 \\
         \bottomrule
        \end{tabular}
\end{table}

\subsection{PAPR}

\begin{figure}
    \centering
    \includegraphics[width=0.42\textwidth]{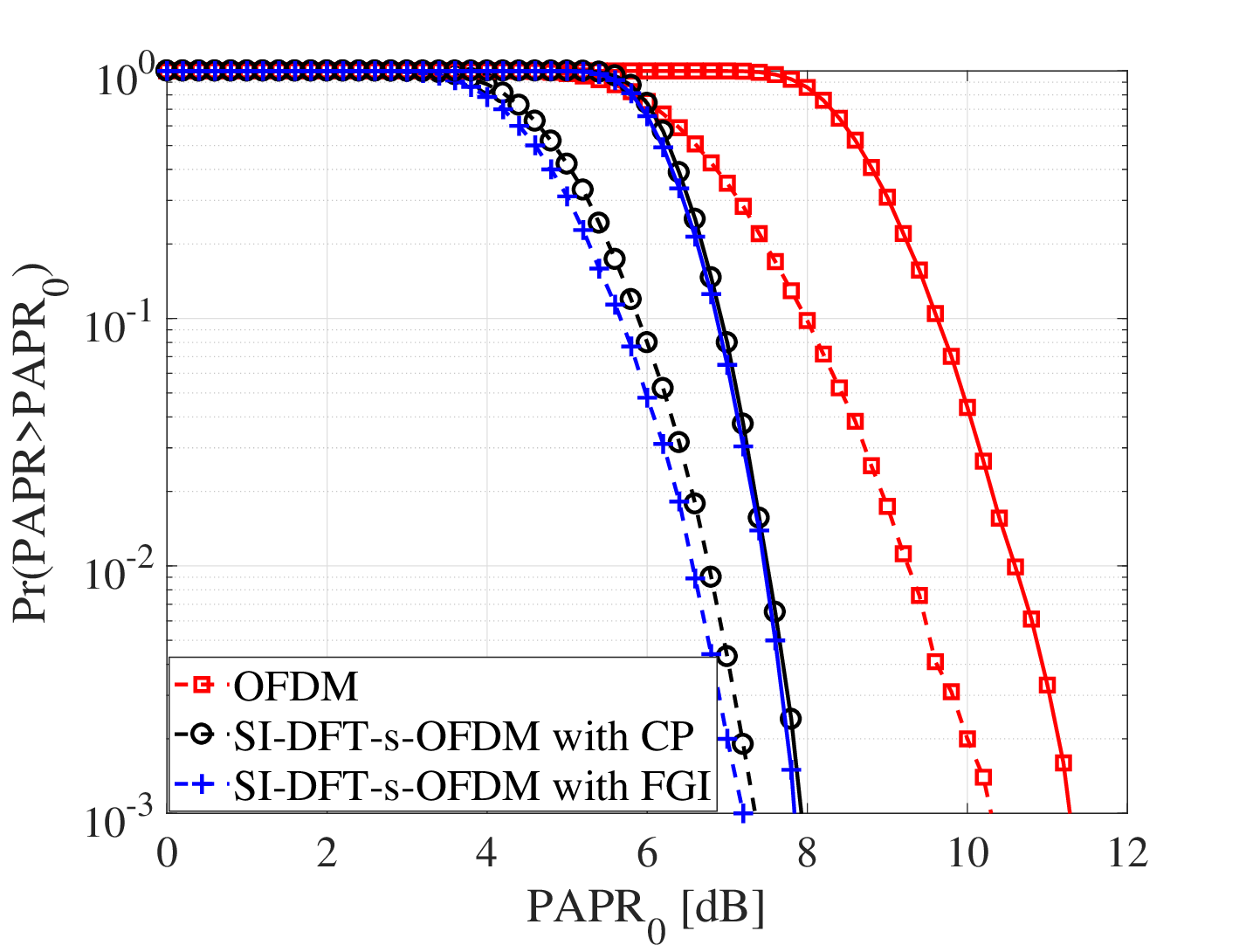}
    \caption{Comparison of PAPR between OFDM and SI-DFT-s-OFDM, $L = \frac{1}{2} N$, $K_p = \frac{1}{4} L $, the subcarrier number is 64 for dotted line and 1024 for solid line.}
    \label{fig:waveform_papr_ccdf}
\end{figure}

\begin{figure}
        \centering
    \includegraphics[width=0.42\textwidth]{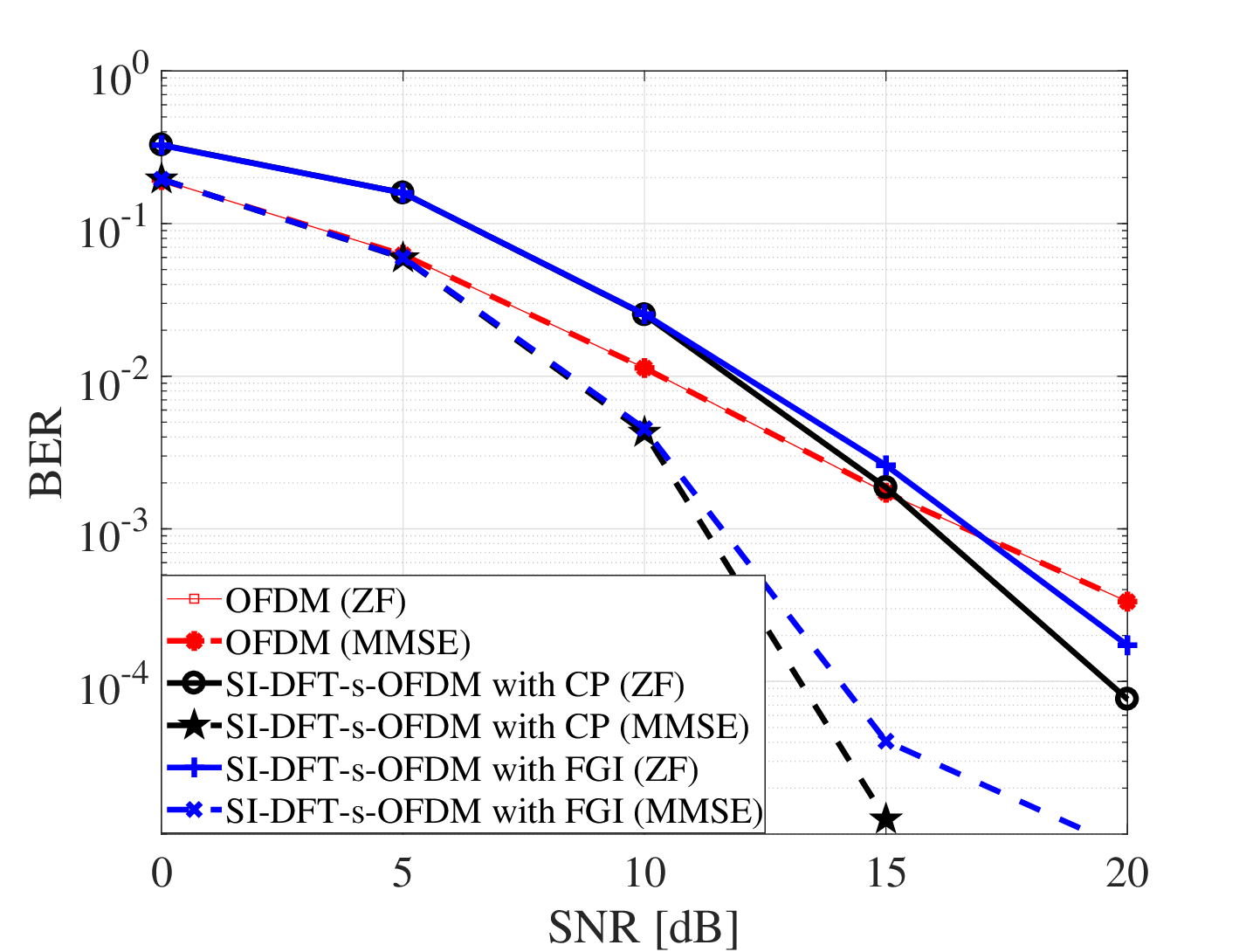}
    \caption{Comparison of BER performance between OFDM and SI-DFT-s-OFDM.
    }
    \label{fig:ber_expert}
\end{figure}

The PAPR of the transmit signal block is a significant characteristic of the waveform in the THz band defined as
\begin{equation}
    \text{PAPR}\{\mathbf{x}_m\} = \frac{\max_{n} |x_{m, n}|^2}{\mathbb{E}\{|x_{m, n}|^2\}}.
\end{equation}
In Fig.~\ref{fig:waveform_papr_ccdf}, we evaluate the PAPR of the SI-DFT-s-OFDM and OFDM signals. The performance metric is the complementary cumulative distribution function (CCDF) of PAPR, i.e., $\text{Pr}(\text{PAPR}>\text{PAPR}_0)$.
We learn that the SI-DFT-s-OFDM has lower PAPR than OFDM for both cases of CP and FGI. In particular, the PAPR values of SI-DFT-s-OFDM data block are approximately 2.6~dB and 3.2~dB lower than OFDM at the CCDF of 1\%, when the subcarrier number is 64 and 1024, respectively. In addition, the PAPR of SI-DFT-s-OFDM with FGI is slightly lower than that with CP. By reducing PAPR, the power backoff of PA can be decreased and the transmit power can be maximized when the saturated power of PA is fixed. Thus, SI-DFT-s-OFDM is able to provide higher coverage and promote more energy-efficient THz communication and sensing than OFDM.

\subsection{Waveform Comparison}

\begin{figure}
    \centering
    \includegraphics[width=0.42\textwidth]{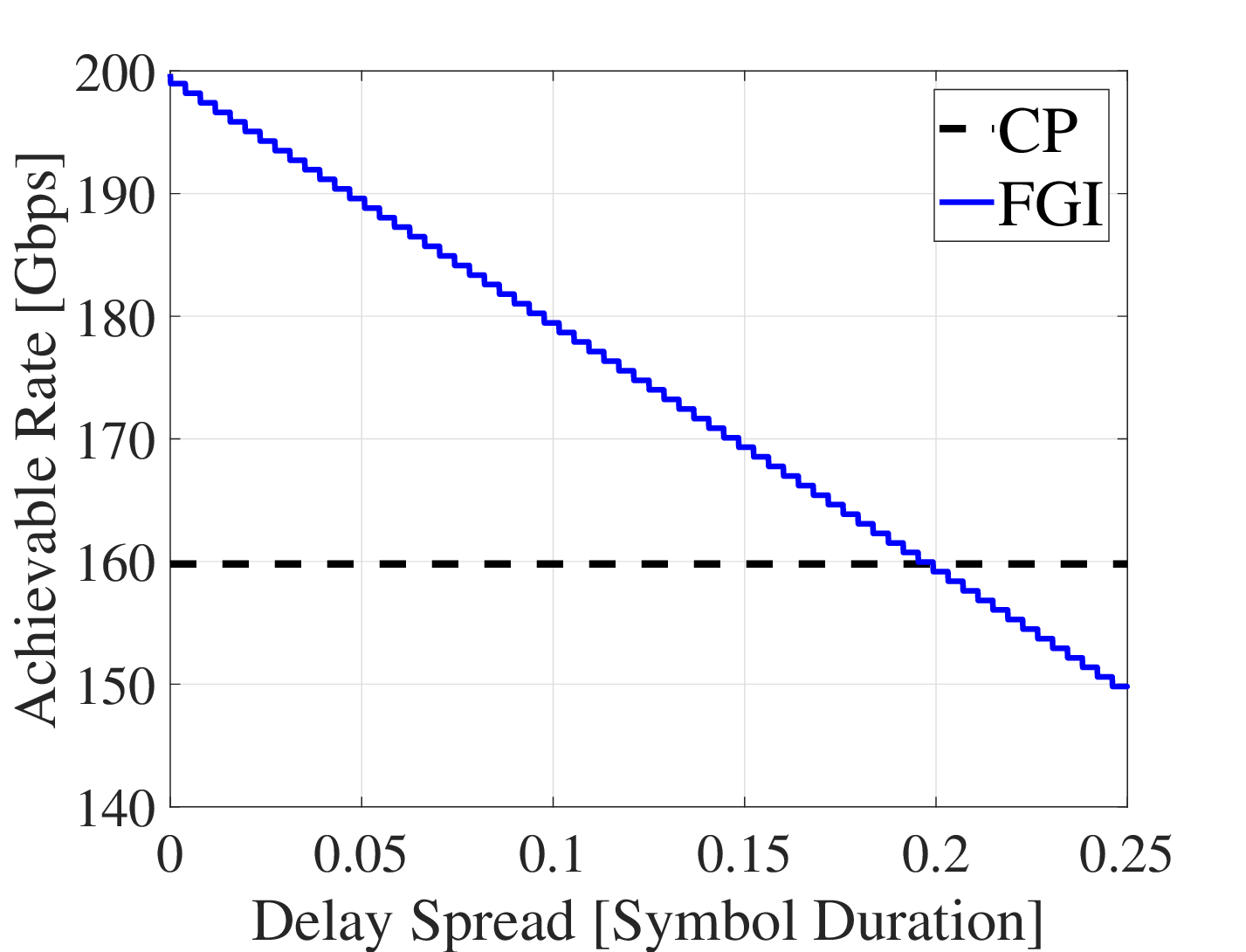}
    \caption{Comparison of achievable rate versus channel delay spread between using CP and FGI, SNR = 20 dB, bandwidth = 30 GHz.}
    \label{fig:achievable_rate}
\end{figure}

We further compare the BER performance of SI-DFT-s-OFDM and OFDM. In our simulation, the number of NLoS paths is set to 4 and the reflection loss in dB unit is assumed to be a Gaussian random variable with the mean -13~dB and the standard deviation 2~dB~\cite{wu2020interference}. The block size and the number of subcarriers are 128 and 256, respectively.

In Fig.~\ref{fig:ber_expert}, we perform both ZF and MMSE equalization for SI-DFT-s-OFDM and OFDM. We learn that with the ZF equalization, the SI-DFT-s-OFDM has higher BER than OFDM below the SNR of 15 dB. However, at high SNR regime, the SI-DFT-s-OFDM can achieve better BER performance for both two equalization methods. In particular, when using the MMSE equalization, the SI-DFT-s-OFDM can improve more than 5 dB gain at the 10\textsuperscript{-3} BER level compared to the OFDM.
The ZF equalizer can amplify the influence of the white noise, especially through the channels with strong frequency-selectivity. Nevertheless, the reflection loss in the THz band results in strong power losses of NLoS paths and reduces the frequency-selectivity of the THz channels. Meanwhile, data symbols are directly modulated on the subcarriers in OFDM and the frequency domain signal has a constant amplitude when using 4-QAM modulation scheme. Therefore, the ZF equalization and MMSE equalization have the same BER performance for OFDM through the THz channels. In the SI-DFT-s-OFDM system, the amplitudes of the frequency domain signal vary greatly and are smaller than the white noise at some subcarriers. In this case, the MMSE method can reduce the influence of white noise on SI-DFT-s-OFDM.

In addition, we calculate the achievable rate of using different guard interval schemes, i.e., CP and FGI. The CP duration is fixed as $\frac{1}{4}T$ and the FGI duration is adjusted according to the channel delay spread, which does not require the adjustment of the waveform numerology. In Fig.~\ref{fig:achievable_rate}, it is indicated that the FGI scheme can support higher achievable rates of 30 Gbps than the CP scheme when the delay spread is 5\% of the symbol duration, by reducing the overhead of the guard interval. The mean value of the achievable rate using the FGI scheme is 174 Gbps, which is more than that using the CP scheme by approximately 14 Gbps. Since the channel sparsity in the THz band may lead to small delay spread in many cases, SI-DFT-s-OFDM with FGI is more promising than that with CP for THz communications.

\subsection{DL-based Sensing Parameter Estimation}

\begin{figure}[t]
    \centering
    \subfigure[P = 1.]{\includegraphics[width=0.42\textwidth]{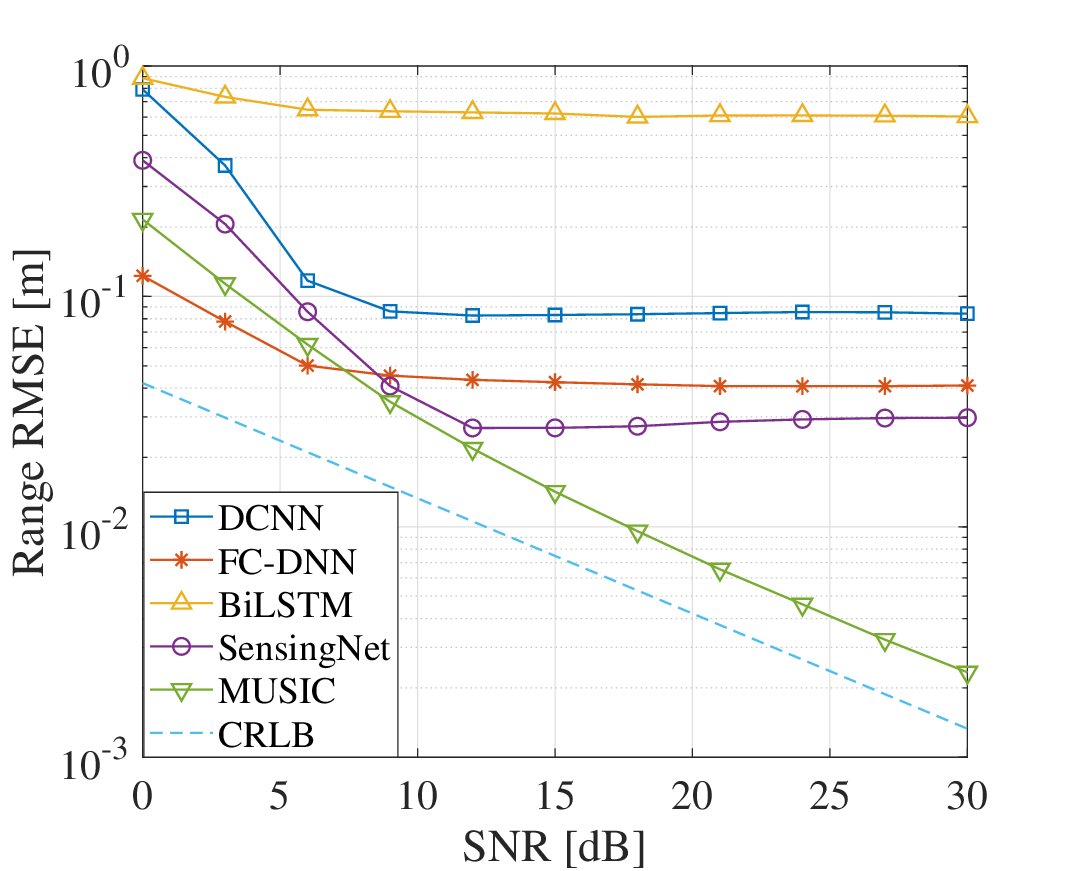}\label{fig:range_rmse_1}}
    \subfigure[P = 2.]{\includegraphics[width=0.42\textwidth]{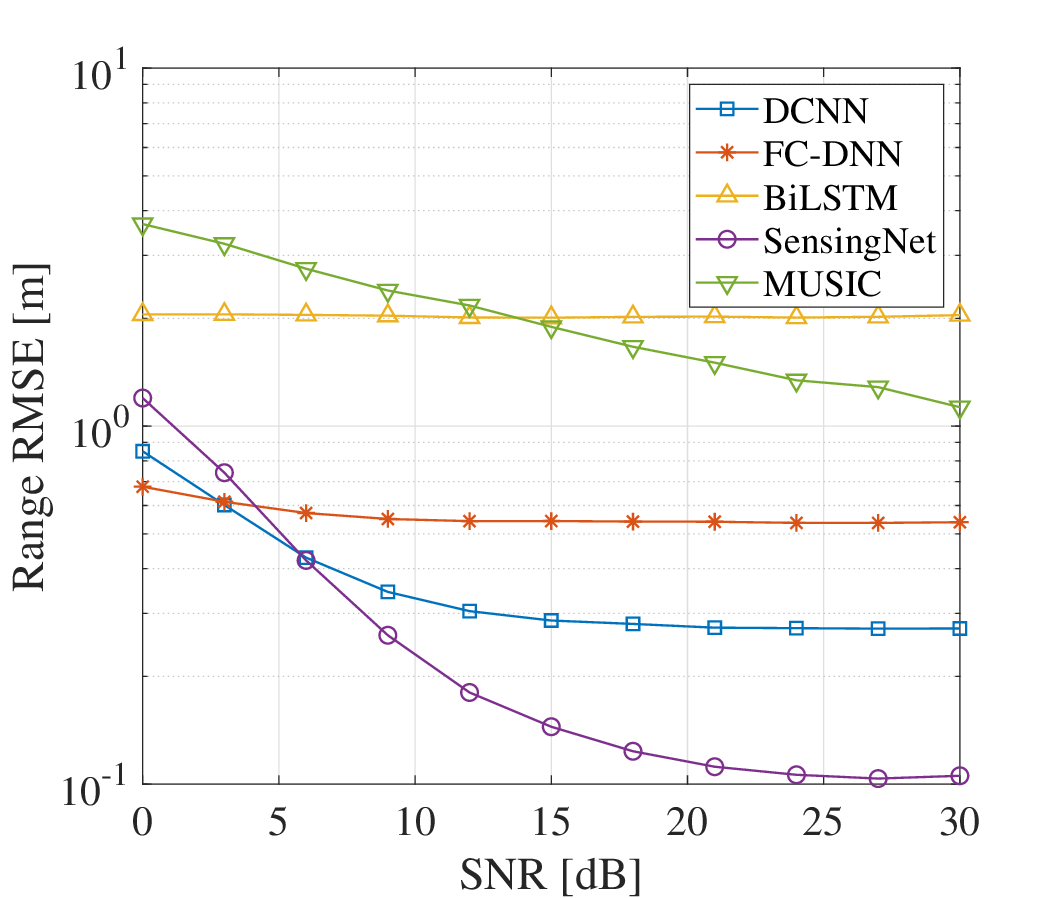}\label{fig:range_rmse_2}}
    \caption{Comparison of range estimation accuracy using different methods.}
    \label{fig:range_rmse}
\end{figure}

Furthermore, we investigate the performance of a single-task DL-based receiver for active sensing, which is implemented by the proposed SensingNet network. In Fig.~\ref{fig:range_rmse}, we compare the range estimation accuracy as a function of SNR, based on the deep learning methods and the MUSIC algorithm. For sensing parameter estimation, we compare the proposed SensingNet with bi-directional long short-term memory (BiLSTM) network~\cite{fan2021thz}, deep convolutional neural network (DCNN)~\cite{chen2020} and fully-connected deep neural network (FC-DNN). The target distances are randomly generated between 0 and $\frac{c_0 T_\text{cp}}{2}$. The subcarrier spacing is set as 7.68 MHz and the block size equals to 32. The learning rate is set as 0.001 when training neural networks.

The simulation results indicate that the RMSE of the range estimation can achieve below 10\textsuperscript{-2} m, i.e., millimeter-level sensing accuracy. As shown in Fig.~\ref{fig:range_rmse_1}, when single target is estimated, the estimation accuracy of the proposed SensingNet is higher than other deep learning methods, while MUSIC performs better than the DL methods at high SNRs. When considering multiple targets, we observe that the MUSIC algorithm requires that different targets are resolvable, i.e., the distance difference among the targets is larger than the resolution of MUSIC. If this condition is not satisfied, MUSIC is not able to distinguish two targets and hence, estimate their distances incorrectly. In Fig.~\ref{fig:range_rmse_2}, when the target number $P$ equals to 2, the DL method achieves better range resolution than MUSIC and is more robust to multi-target estimation.

\begin{figure}
    \centering
    \includegraphics[width=0.42\textwidth]{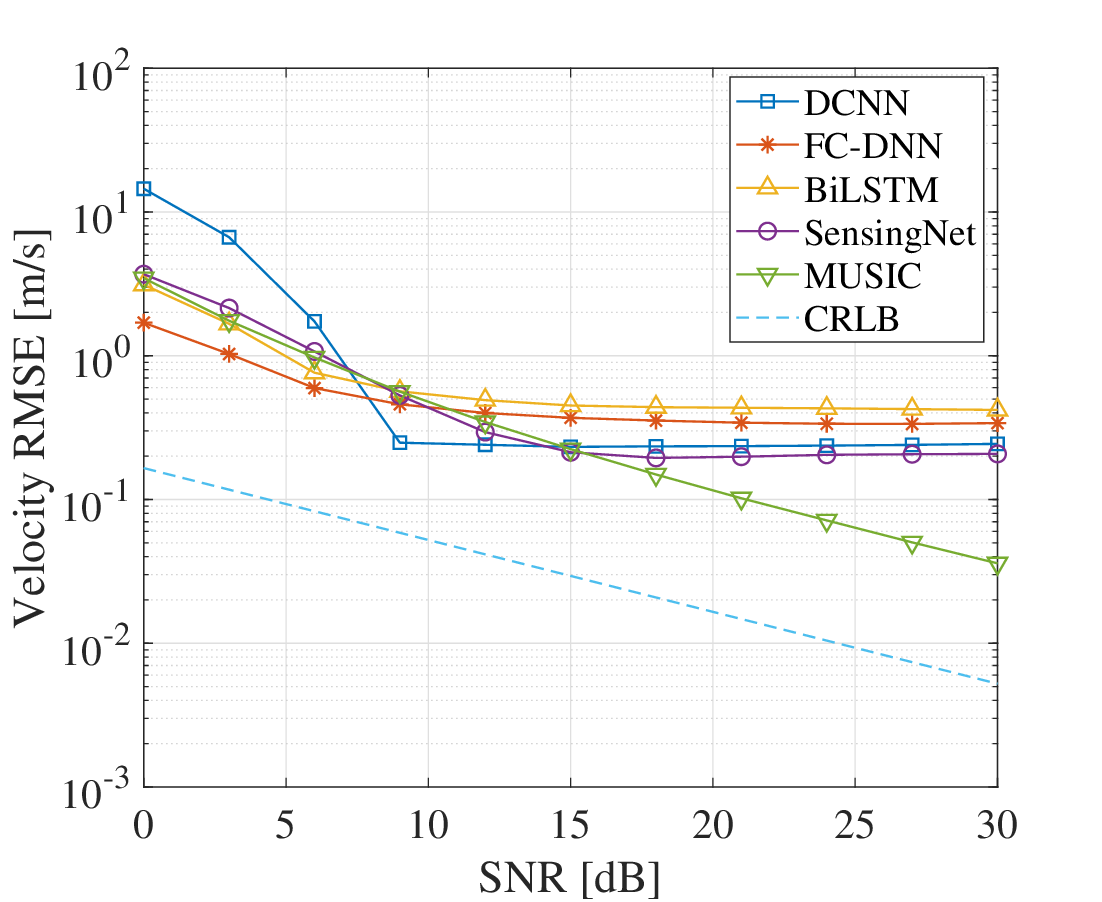}
    \caption{Comparison of velocity estimation accuracy using different methods.}
    \label{fig:velocity_rmse_1}   
\end{figure}

Moreover, the velocity estimation accuracy using different methods is compared in Fig.~\ref{fig:velocity_rmse_1}, in which the target velocity is randomly generated between -100 km/h and 100 km/h, and the number of reference blocks equals to 32. The simulation results indicate that velocity estimation with the proposed SI-DFT-s-OFDM is able to achieve the decimeter-per-second level accuracy. Among the concerned DL methods, our proposed SensingNet network achieves the best velocity estimation accuracy.

\begin{figure}[t]
    \centering
    \subfigure[BER performance of SI-DFT-s-OFDM with CP.]{\includegraphics[width=0.42\textwidth]{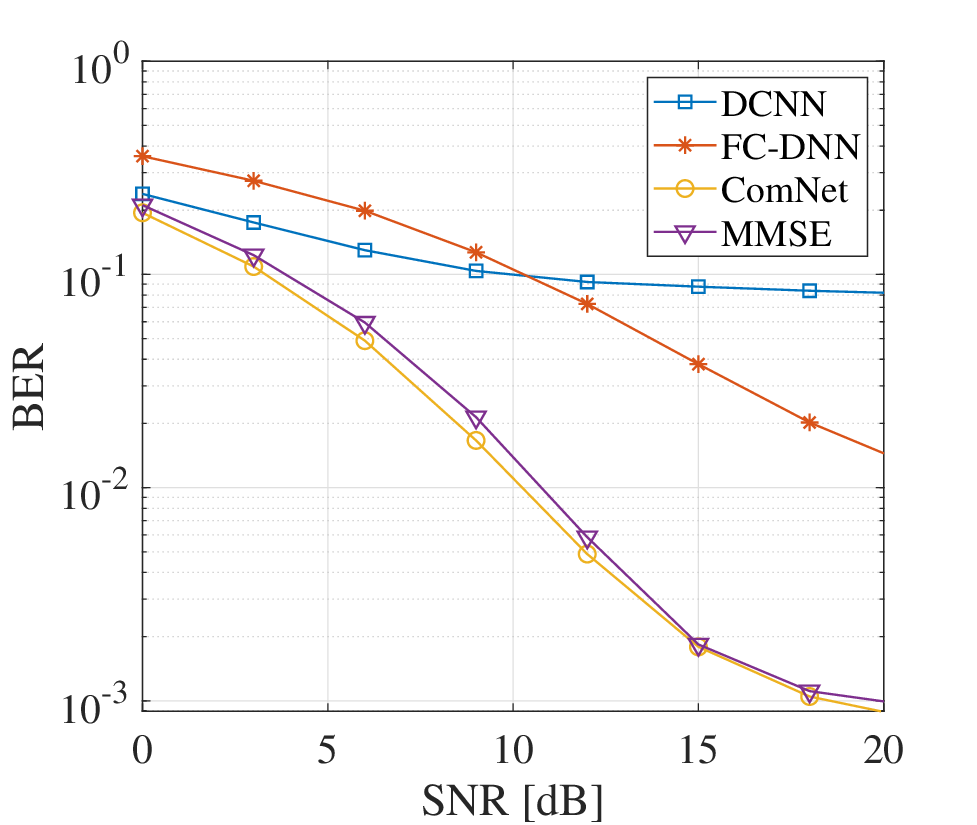}\label{fig:dftsofdm_doppler_ber}}
    \subfigure[BER performance of SI-DFT-s-OFDM with FGI.]{\includegraphics[width=0.42\textwidth]{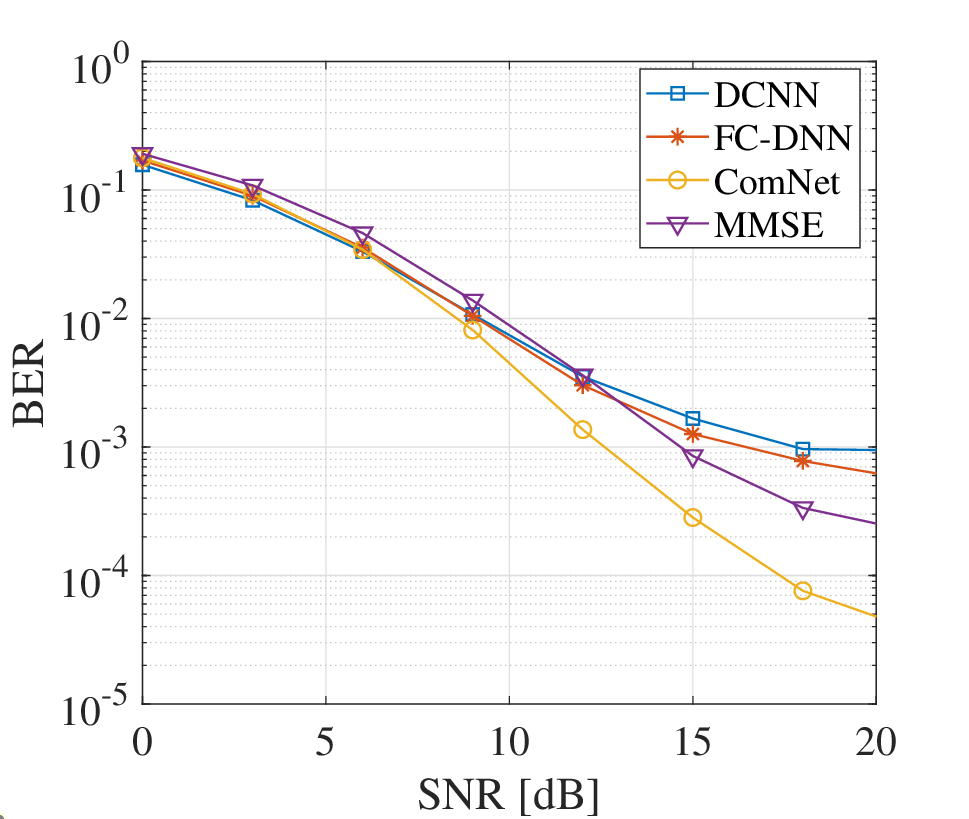}\label{fig:fgi_doppler_ber}}
    \caption{Comparison of BER performance using different methods under Doppler effects.}
    \label{fig:doppler_ber}
\end{figure}

\begin{figure}
    \centering
    \includegraphics[width=0.42\textwidth]{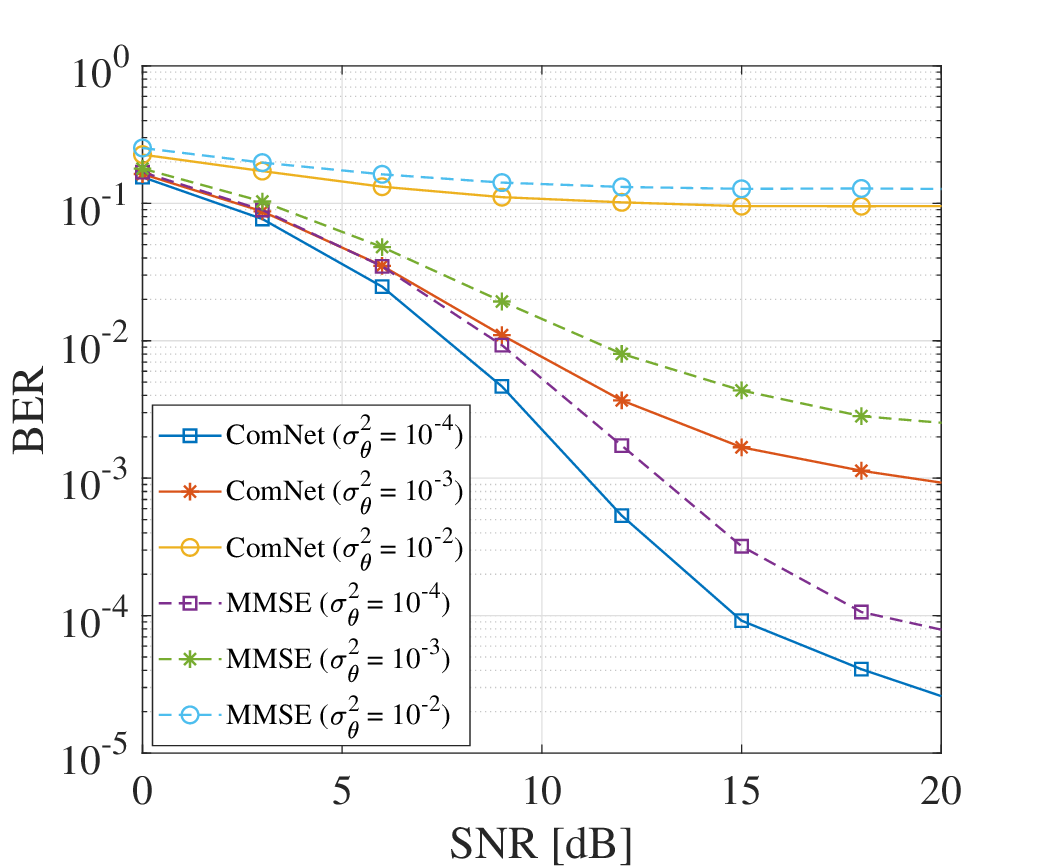}
    \caption{BER performance of SI-DFT-s-OFDM with FGI using ComNet and MMSE under phase noise effects.}
    \label{fig:fgi_pn_ber}
\end{figure}

\subsection{DL-based Data Detection}

Next, we conduct the performance evaluation of the DL-based communication receiver with the proposed ComNet network under Doppler and phase noise effects. In addition to the classical MMSE equalization method, we compare the ComNet network with DCNN~\cite{chen2020} and FC-DNN~\cite{ye2019dl} in terms of the BER performance. The subcarrier spacing is set as 1.92 MHz, and the interval of the reference blocks in a frame $S_r$ equals to 10. The speed along each path is randomly generated between -100 km/h and 100 km/h. In Fig.~\ref{fig:dftsofdm_doppler_ber} and Fig.~\ref{fig:fgi_doppler_ber}, the proposed two-level ComNet method outperforms MMSE, DCNN and FC-DNN. Specifically, when using FGI scheme, the proposed ComNet is able to improve 2 dB performance gain at the 10\textsuperscript{-3} BER level compared to the MMSE equalization.

\begin{table*}[t]
    \centering
    \caption{Comparison of Computational Complexity and Running Time for Different Methods}
    \begin{tabular}{cccc}\toprule
        Method & Computational Complexity & Running Time for Sensing (ms) & Running Time for Communication (ms) \\
        \midrule
        MUSIC & $\mathcal{O}(M_\text{RB} L^2), \mathcal{O}(M_\text{RB}^2 L)$ & 1.2 & - \\
        FC-DNN & $\mathcal{O}(M_\text{RB} L), \mathcal{O}(M_\text{DB} L) $ & 0.08 & 0.43 \\
        BiLSTM & $\mathcal{O}(M_\text{RB} L) $ & 3.1 & - \\
        DCNN & $\mathcal{O}(M_\text{RB} L), \mathcal{O}(M_\text{DB} L) $ & 0.13 & 0.13 \\
        SensingNet & $\mathcal{O}(M_\text{RB} L) $ & 0.58 & - \\
        MMSE & $\mathcal{O}(M_\text{DB} L) $ & - & 0.59 \\
        ComNet & $\mathcal{O}(M_\text{DB} L) $ & - & 1.1 \\
    \bottomrule\\
    \end{tabular}
    \label{tab:complexity}
\end{table*}

Then, the influence of the phase noise in the THz band is studied. We set the phase phase noise parameter $\sigma_\theta^2$ as 10\textsuperscript{-4}, 10\textsuperscript{-3} and 10\textsuperscript{-2}. We evaluate the BER performance of SI-DFT-s-OFDM with FGI in presence of phase noise by using the ComNet network and the MMSE method. Since ComNet outperforms MMSE under these phase noise effects, we state that the proposed DL method has stronger robustness to phase noise.

The computational complexity and running time for each sample of the proposed DL methods and literature solutions are compared in Table~\ref{tab:complexity}. We observe that the proposed SensingNet has lower computational complexity and running time than MUSIC. While the FC-DNN and DCNN process less running time than SensingNet, they have higher estimation error. Apart from that, the proposed ComNet spends more running time than MMSE but can achieve better BER performance under Doppler and phase noise effects.

While the user mobility can cause the variations of the wireless environment, it is not necessary to retrain the proposed DL model when the user moves. In our simulations, we consider generating each sample cased on a stochastic channel with random parameters. The proposed DL model is robust against the variations of channel parameters, since the pilot signal in the input features of the neural network can contribute to symbol detection and parameter estimation regardless of the channel variations. In real-world applications, while mismatches of channel statistics may occur between the offline training stage and online deployment, variations on statistics of channel models do not have significant damage on the performance of symbol detection~\cite{ye2019dl}. If the channel model type, rather than the parameters or statistics, changes due to the user mobility, which is rare in practice, one needs to apply different DL models that are trained in the offline stage. Thus, the duration of training process does not cause a problem in practical applications.

\begin{figure}
     \centering
    \includegraphics[width=0.42\textwidth]{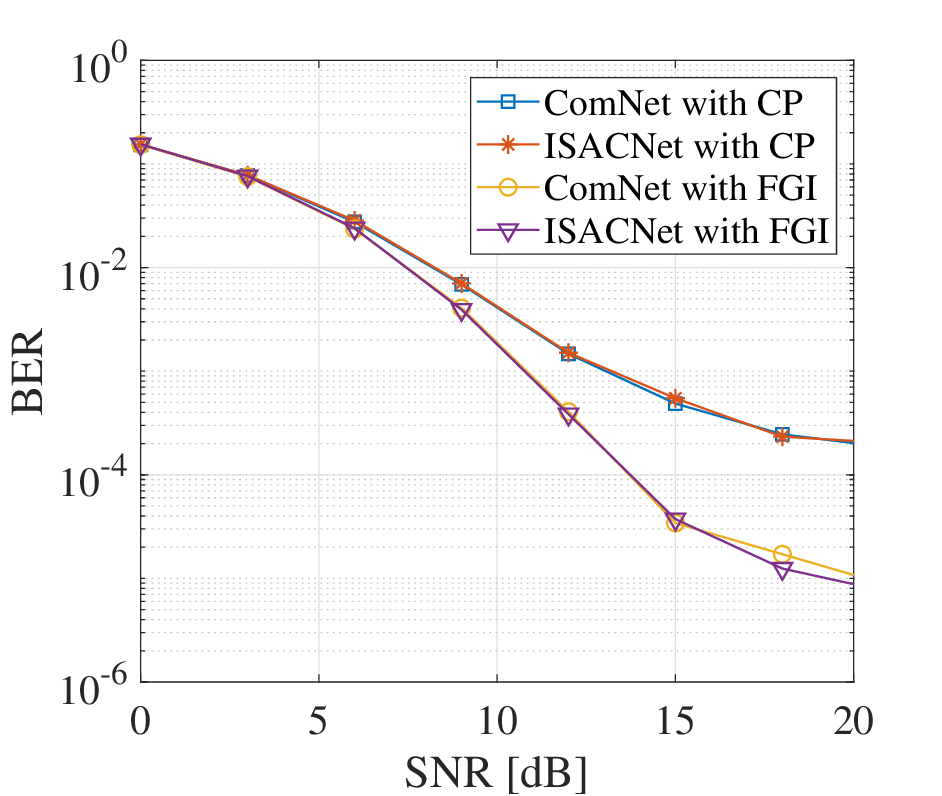}
    \caption{Comparison of BER performance using ISACNet and separate models.}
    \label{fig:mtl_ber} 
\end{figure}

\begin{figure}
     \centering
    \includegraphics[width=0.42\textwidth]{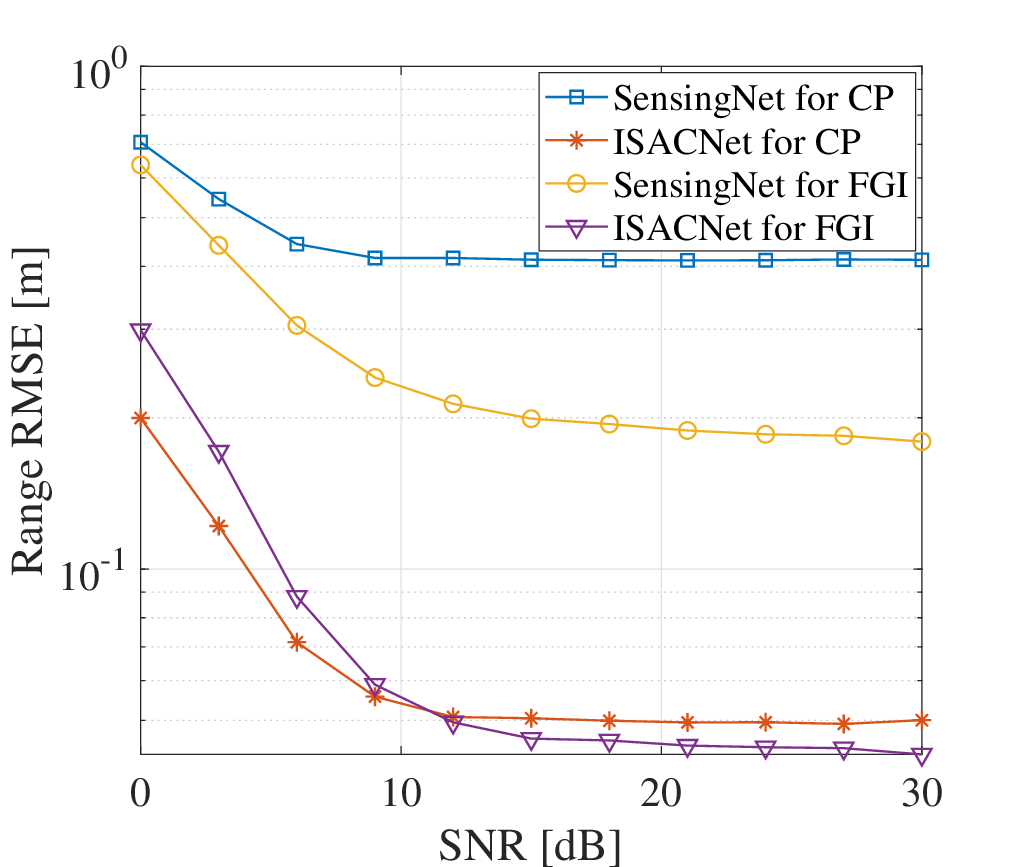}
    \caption{Comparison of range estimation performance using ISACNet and separate models.}
    \label{fig:mtl_rmse} 
\end{figure}

\subsection{Two-Task DL-based Receiver for Passive Sensing}

Finally, we investigate the performance of two-task DL-based ISAC receiver for passive sensing. Herein, the sensing task is to estimate the length of the LoS path. The performance of ISACNet model is compared in Fig.~\ref{fig:mtl_ber} and Fig.~\ref{fig:mtl_rmse} with separate single-task models. The ISACNet network is implemented by sharing 2 dense layers between ComNet and SensingNet. In Fig.~\ref{fig:mtl_ber}, the ISACNet model and the ComNet realize close BER performance. As shown in Fig.~\ref{fig:mtl_rmse}, the range estimation accuracy of ISACNet achieves higher range estimation accuracy than that of separate SensingNet models. In addition, the SensingNet and ComNet have 197179 and 230222 trainable parameters, namely, 427401 parameters in total, while the ISACNet model has 268023 trainable parameters. Thus, in contrast with separate models, the ISACNet model has a reduced network complexity and is able to improve the range estimation accuracy, while maintaining the same BER performance. We calculate the task relatedness of communication and sensing by setting $\eta =1$ in \eqref{eq:relatedness}, which equals to 0.9643 for the CP scheme and 0.8555 for the FGI scheme. The task relatedness for the CP scheme is higher than that for the FGI scheme, i.e., sensing and communication are more highly related for the CP scheme than that for the FGI scheme. Thus, in Fig.~\ref{fig:mtl_ber}, the performance gain for CP is greater than that for FGI.

\section{Conclusion}\label{sec:conclusion}

In this paper, we have proposed a sensing integrated DFT-s-OFDM system for THz ISAC. We design two types of THz waveforms, i.e., CP based SI-DFT-s-OFDM and FGI based SI-DFT-s-OFDM, which utilize the specific features of THz channels and take into account the requirements of THz transceivers. Furthermore, we have developed a deep learning powered receiver to simultaneously perform sensing parameter estimation and signal recovery.

With extensive simulation, the results have demonstrated that the proposed SI-DFT-s-OFDM can reduce the PAPR by approximately 3.2~dB and enhance 5~dB gain at the 10\textsuperscript{-3} BER level in the THz channel, compared to the OFDM system. The proposed SI-DFT-s-OFDM with the FGI scheme can achieve a mean achievable rate of 174 Gbps and 10\textsuperscript{-5} BER performance at the SNR of 20 dB, while realizing millimeter-level range estimation and decimeter-per-second-level velocity estimation accuracy. In contrast with the conventional ISAC systems and other DL methods, the proposed DL-based ISAC receiver is more robust to Doppler effects, phase noise and multi-path propagation, which is preferred in THz systems. In particular, the SensingNet method achieves higher accuracy for multi-target estimation and the ComNet performs better BER performance than the classical MMSE equalizer under Doppler and PN effects. Meanwhile, the ISACNet model is able to reduce the network complexity and improve the range estimation accuracy in contrast with separate models.
In this work, it is shown that our designs are effective at low-THz scenarios. At higher THz frequencies, the saturated output value of power amplifiers becomes even lower, the Doppler shifts worsens and THz wave propagation is further attenuated by the atmospheric effects. In future work, To address these more stringent challenges, the performance of the developed waveform needs to be further improved and more robust against these effects.

\bibliographystyle{IEEEtran}
\bibliography{journal}

\begin{IEEEbiography}[{\includegraphics[width=1in,height=1.25in,clip,keepaspectratio]{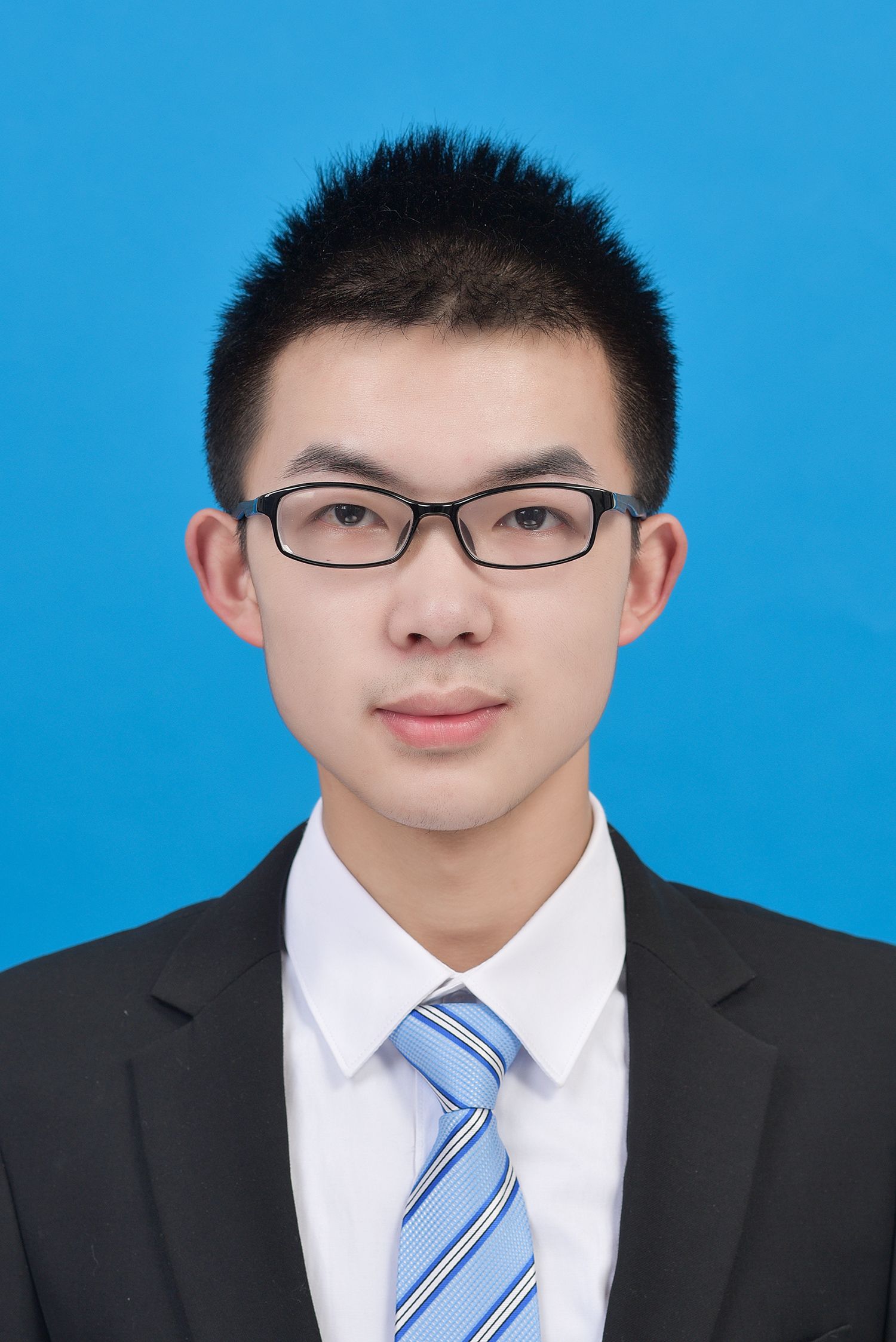}}]{Yongzhi Wu}
(S'19) received B.E degree in Electronic and Information Engineering from Huazhong University of Science and Technology in 2019. Since 2019, he is pursuing Ph.D. degree in the Terahertz Wireless Communication Laboratory, Shanghai Jiao Tong University. His research interests include Terahertz communications, integrated sensing and communication.
\end{IEEEbiography}

\begin{IEEEbiography}[{\includegraphics[width=1in,height=1.25in,clip,keepaspectratio]{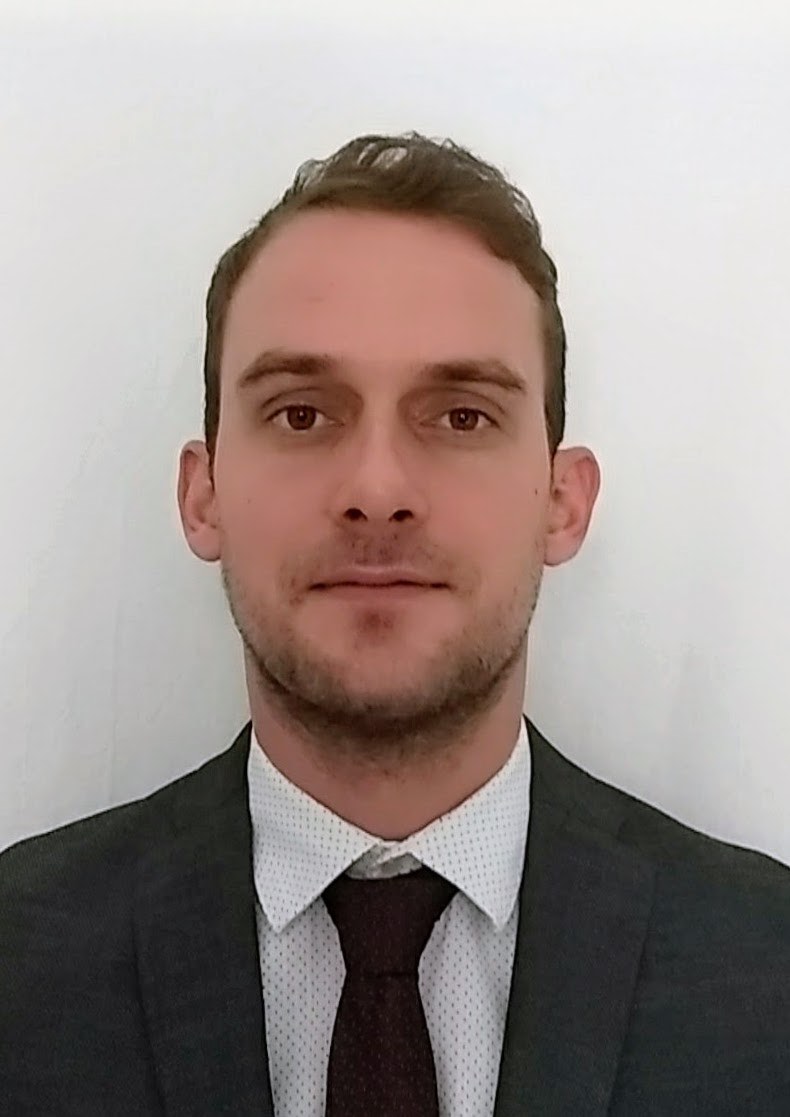}}]{Filip Lemic}
received his B.Sc. and M.Sc. from the University of Zagreb in 2010 and 2012, and his Ph.D. from the Technische Universität Berlin in 2017. Currently, he is a senior researcher at the i2Cat Foundation. He was a postdoctoral researcher and Marie Curie fellow at the University of Antwerp (2018-22) and Universitat Politècnica de Catalunya (2020-22). He was also affiliated with imec (2018-22), FIWARE Foundation (2018), and Technische Universität Berlin (2012-17). He was a visiting researcher at the University of California at Berkeley (2015-16) and Shanghai Jiao Tong University (2019-20). He co-authored more than 60 peer-reviewed research articles and was involved in various international research projects, notably EU MSCA ScaLeITN, EU EVARILOS, EU eWine, NIST’s PerfLoc, and UC Berkeley’s beyond-5G.  
\end{IEEEbiography}

\begin{IEEEbiography}[{\includegraphics[width=1in,height=1.25in,clip,keepaspectratio]{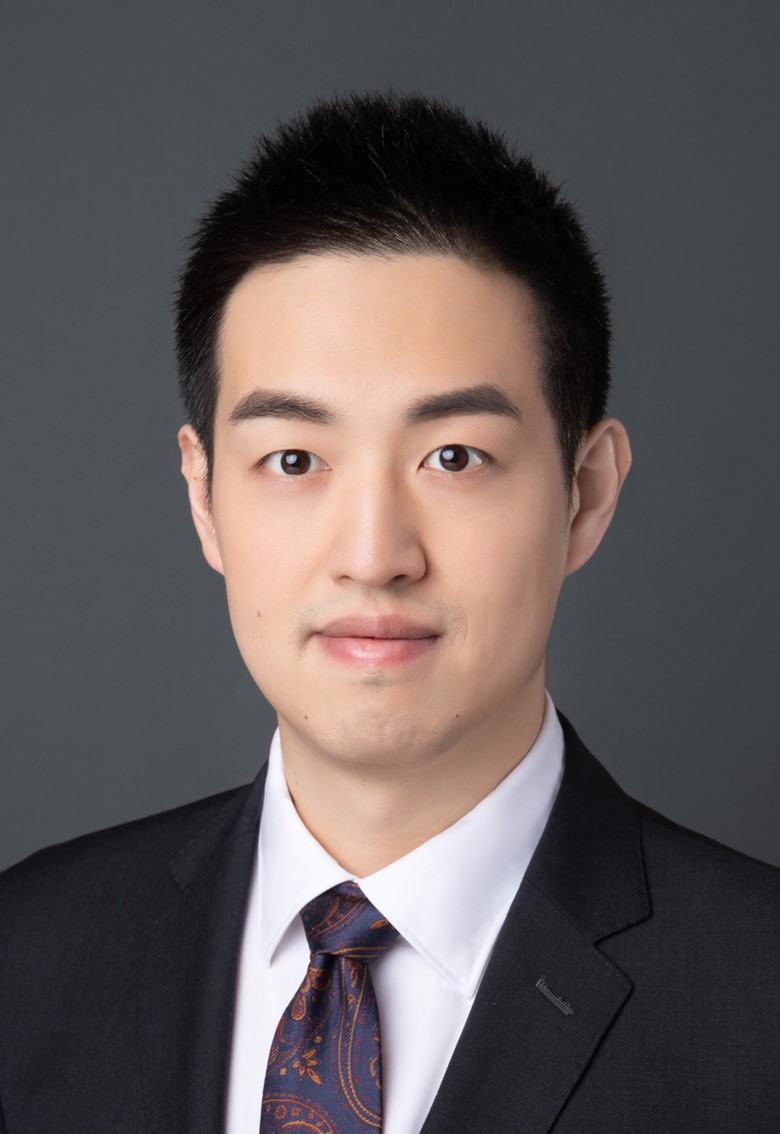}}]{Chong Han}
(M’16) received Ph.D. degree in Electrical and Computer Engineering from Georgia Institute of Technology, USA in 2016. He is currently an John Wu \& Jane Sun Endowed Associate Professor with University of Michigan-Shanghai Jiao Tong University (UM-SJTU) Joint Institute, Shanghai Jiao Tong University, China, and director of the Terahertz Wireless Communications (TWC) Laboratory. Since 2021, he is also affiliated with Department of Electronic Engineering, Shanghai Jiao Tong University. He is the recipient of 2018 Elsevier NanoComNet (\textit{Nano Communication Network Journal}) Young Investigator Award, 2017 Shanghai Sailing Program 2017, and 2018 Shanghai ChenGuang Program. He is a guest editor with \textsc{IEEE Journal on Selected Topics in Signal Processing (JSTSP)} and \textsc{IEEE Transactions on Nanotechnology}, an editor with \textsc{IEEE Open Journal of Vehicular Technology} since 2020, \textsc{IEEE Access} since 2017, \textit{Elsevier Nano Communication Network Journal} since 2016. He is a TPC chair to organize multiple IEEE and ACM conferences and workshops. He is a co-founder and vice-chair of IEEE ComSoc Special Interest Group (SIG) on Terahertz Communications, since 2021. His research interests include Terahertz and millimeter-wave communications. He is a member of the IEEE and ACM.
\end{IEEEbiography}

\begin{IEEEbiography}[{\includegraphics[width=1in,height=1.25in,clip,keepaspectratio]{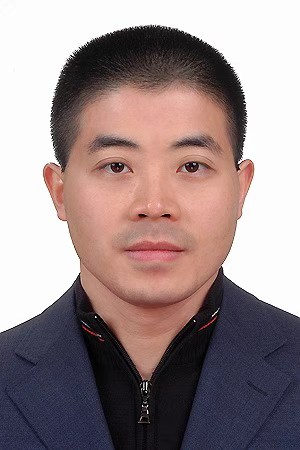}}]{Zhi Chen}
(SM'16) received B. Eng, M. Eng., and Ph.D. degree in Electrical Engineering from University of Electronic Science and Technology of China (UESTC), in 1997, 2000, 2006, respectively. On April 2006, he joined the National Key Lab of Science and Technology on Communications (NCL), UESTC, and worked as a professor in this lab from August 2013. He was a visiting scholar at University of California, Riverside during 2010-2011. He is also the deputy director of Key Laboratory of Terahertz Technology, Ministry of Education. His current research interests include Terahertz communication, 5G mobile communications and tactile internet. 
\end{IEEEbiography}

% \ifCLASSOPTIONcaptionsoff
%   \newpage
% \fi

\end{document}